\documentclass[a4paper,12pt]{article}

\usepackage[T1]{fontenc}
\usepackage[utf8]{inputenc}
\usepackage{lmodern}
\usepackage{fancyhdr}
\usepackage[margin = 2.5cm]{geometry}
\usepackage{setspace}
\usepackage{amsmath}
\usepackage{amssymb}
\usepackage{pdfpages}
\usepackage{mathabx, stmaryrd}
\usepackage{mathrsfs}
\usepackage{tikz,pgf}
\usetikzlibrary{arrows.meta}
\usepackage{indentfirst}
\usepackage{dutchcal} 
\usepackage{float}
\usepackage[centerlast,small,sc]{caption}
\usepackage{subcaption}\usepackage[indent = 20pt, skip = .5\baselineskip plus 1pt]{parskip}%
\usepackage{lastpage}
\usepackage[bottom]{footmisc}
\usepackage[unicode, pdfstartview = FitH, colorlinks, linkcolor = blue]{hyperref}
\usepackage{babel}
\usepackage{listings}
\usepackage[linesnumbered, ruled, vlined,onelanguage]{algorithm2e}
\usepackage{datetime}
\usepackage{changepage}

\makeatletter%
	\newcommand{\tableofcontentsAnnex}{\@starttoc{tocannex}}%
\makeatother%

\SetKw{Continue}{continue}

\begin{document}

	\renewcommand*{\labelitemi}{--}%
	\renewcommand*{\labelitemii}{$\bullet$}%
	\renewcommand*{\labelenumi}{{\textbf{\emph{\arabic{enumi}})}}}%
	\renewcommand*{\labelenumii}{{\alph{enumii})}}%
 \renewcommand*{\abstract}{\textbf{Abstract.}\footnotesize}
\newdateformat{monthyeardate}{\monthname[\THEMONTH], \THEYEAR}
 
	\definecolor{bg}{rgb}{0.95,0.95,0.95}

	\SetKwComment{Comment}{/* }{ */}%
    \SetKwRepeat{Do}{do}{while}

    \fancypagestyle{testpage}{%
	\fancyhf{}

	\renewcommand{\footrulewidth}{0.4pt}%
	\renewcommand{\headrulewidth}{0pt}%

	\fancyfoot[L]{\scriptsize \textup{Constrained and $k$ Shortest Paths}}%
	\fancyfoot[C]{-- \thepage{} --}%
	\fancyfoot[R]{\scriptsize A. BENDAHI, A. FRADIN}%
    }


\renewcommand*{\thesection}{\Roman{section}}
\renewcommand*{\thesubsection}{\arabic{subsection})}
\renewcommand*{\thesubsubsection}{\roman{subsubsection})}

\newcommand{\commentalgo}[1]{\footnotesize\ttfamily\textcolor{blue}{#1}}%
\SetCommentSty{commentalgo}%
\SetKwComment{Comment}{/* }{ */}%
\SetKwProg{Init}{init}{}{}

\newcommand*{\centrer}[1]{\par\hfil#1\hfil\par}%


\newcommand*{\lbd}{\lambda}
\newcommand*{\eps}{\varepsilon}

\newcommand*{\zeros}{\mathop{\mathscr{Z}}}
\renewcommand*{\deg}{\mathop{\mathrm{deg}}}
\newcommand*{\dom}{\mathop{\mathrm{dom}}}
\newcommand*{\mul}{\mathop{\mathrm{m}}}
\newcommand*{\res}{\mathop{\mathrm{Res}}}

\newcommand*{\E}[1]{\left\lfloor #1 \right\rfloor}
\newcommand*{\Es}[1]{\left\lceil #1 \right\rceil}
\newcommand*{\val}{\nu}
\renewcommand*{\divides}{\mid}%
\renewcommand*{\notdivides}{\nmid}%

\newcommand*{\pgcd}{\mathop{\mathrm{PGCD}}}
\newcommand*{\ppcm}{\mathop{\mathrm{PPCM}}}

\newcommand*{\primepi}{\mathop{\pi}}

\newcommand*{\mat}{\mathop{\mathrm{mat}}\limits}%
\newcommand*{\coord}{\mathop{\mathrm{Coord}}\limits}%
\newcommand*{\T}[1]{{\vphantom{#1}}^{\mathbf t}{#1}}

\newcommand*{\Ker}{\mathop{\mathrm{Ker}}}
\newcommand*{\Img}{\mathop{\mathrm{Im}}}

\newcommand*{\rg}{\mathop{\mathrm{rg}}}

\newcommand*{\Id}{\mathop{\mathrm{Id}}}
\newcommand*{\Tr}{\mathop{\mathrm{Tr}}}

\renewcommand*{\det}{\mathop{\mathrm{det}}}

\newcommand*{\Com}{\mathop{\mathrm{Com}}}

\newcommand*{\iddots}{\rotatebox[origin = c]{90}{$\ddots$}}

\newcommand*{\GL}{\mathop{\mathrm{GL}}}
\newcommand*{\SL}{\mathop{\mathrm{SL}}}
\renewcommand*{\O}{\mathop{\mathrm{O}}}
\newcommand*{\SO}{\mathop{\mathrm{SO}}}
\newcommand*{\Sto}{\mathop{\mathrm{ST}}}
\newcommand*{\BSto}{\mathop{\mathrm{BST}}}

\newcommand*{\Sp}{\mathop{\mathrm{Sp}}}

\renewcommand*{\dim}{\mathop{\mathrm{dim}}}
\newcommand*{\Vect}[2][]{\mathop{\mathrm{Vect}}_{#1}\left(#2\right)}

\newcommand*{\intervalle}[4]{\mathopen{#1}#2\mathclose{}\mathpunct{};#3\mathclose{#4}}%
\newcommand*{\intff}[2]{\intervalle{[}{#1}{#2}{]}}
\newcommand*{\intof}[2]{\intervalle{]}{#1}{#2}{]}}
\newcommand*{\intfo}[2]{\intervalle{[}{#1}{#2}{[}}
\newcommand*{\intoo}[2]{\intervalle{]}{#1}{#2}{[}}
\newcommand*{\Int}[2]{\intervalle\llbracket{#1}{#2}\rrbracket}

\newcommand*{\ens}[1]{\left\{#1\right\}}%
\newcommand*{\enstq}[2]{\left\{#1\,\middle/\,#2\right\}}%
\newcommand*{\Card}[1]{\left|#1\right|}
\newcommand*{\Partie}{\mathop{\mathcal{P}}}
\newcommand*{\comp}[1]{#1^\mathrm{c}}

\newcommand*{\N}{\mathbb{N}}
\renewcommand*{\P}{\mathbb{P}}
\newcommand*{\Z}{\mathbb{Z}}
\newcommand*{\Q}{\mathbb{Q}}
\newcommand*{\R}{\mathbb{R}}
\newcommand*{\C}{\mathbb{C}}
\renewcommand*{\H}{\mathbb{H}}
\newcommand*{\K}{\mathbb{K}}
\renewcommand*{\L}{\mathbb{L}}
\newcommand*{\F}{\mathbb{F}}
\newcommand*{\U}{\mathbb{U}}
\renewcommand*{\S}{\mathbb{S}}

\newcommand*{\Hol}{\mathrm{Hol}}
\newcommand*{\Mer}{\mathcal{M}}
\newcommand*{\Resi}{\mathrm{Res}}
\newcommand*{\Ind}{\mathrm{Ind}}

\newcommand*{\RelB}{\mathop{\mathscr{R}}}
\newcommand*{\Cl}[1]{\overline{#1}}

\newcommand*{\gr}{\mathop{\mathrm{gr}}}
\newcommand*{\ord}{\mathop{\mathrm{ord}}}

\newcommand*{\Orb}{\mathop{\mathrm{Orb}}}

\newcommand*{\Car}{\mathop{\mathrm{Car}}}


\newcommand*{\Alg}{\mathop{\mathrm{Alg}}} 

\newcommand{\dist}{\mathop{\mathrm{d}}}
\newcommand{\diam}{\delta}
\newcommand{\normesub}[1]{{\left\vert\left\vert\left\vert #1 \right\vert\right\vert\right\vert}}

\newcommand{\bouleO}{\mathop{\mathrm{B}}}
\newcommand{\bouleF}{\overline{\mathop{\mathrm{B}}}}
\newcommand{\sphere}{\mathop{\mathrm{S}}}
\newcommand{\adh}[1]{\overline{#1}}
\newcommand{\inte}[1]{\mathring{#1}}
\newcommand{\ext}{\mathop{\mathrm{ext}}}
\newcommand{\fr}{\mathop{\mathrm{fr}}}

\renewcommand*{\Bar}{\mathop{\mathrm{Bar}}}
\newcommand*{\Conv}{\mathop{\mathrm{Conv}}}

\newcommand{\Epi}{\mathop{\mathrm{Epi}}}

\newcommand{\ortho}[1]{{#1}^{\bot}}
\newcommand{\orthosum}{\mathop{\overset{\bot}{\oplus}}}
\newcommand{\orthoproj}{\mathop{\mathrm{p}}^{\bot}}

\newcommand{\ps}[2]{\left\langle #1 \mid #2 \right\rangle}

\newcommand*{\Sym}{\mathop{\mathfrak{S}}}
\newcommand*{\sig}{\mathop{\varepsilon}}
\newcommand*{\Inv}{\mathop{\mathrm{Inv}}}

\newcommand*{\conj}[1]{\overline{#1}}
\newcommand*{\abs}[1]{\left\lvert #1 \right\rvert}
\renewcommand*{\Re}{\mathop{\mathrm{Re}}}
\renewcommand*{\Im}{\mathop{\mathrm{Im}}}

\renewcommand*{\lim}{\mathop{\mathrm{lim}}\limits}

\renewcommand*{\o}{\mathop{\mathrm{o}}\limits}
\renewcommand*{\O}{\mathop{\mathrm{O}}\limits}
\newcommand*{\eq}{\mathop{\sim}\limits}

\newcommand{\argsinh}{\mathop{\mathrm{argsinh}}}
\newcommand{\argcosh}{\mathop{\mathrm{argcosh}}}
\newcommand{\argtanh}{\mathop{\mathrm{argtanh}}}

\newcommand{\diff}{\mathop{\!\mathrm{d}}}
\newcommand{\dirdiff}{\mathop{\mathrm{D}}}
\newcommand{\pardiff}[2]{{\displaystyle \frac{\partial #1}{\partial #2}}}
\newcommand{\Jac}{\mathop{\mathrm{J}}}

\newcommand*{\proba}{\mathop{\mathbb{P}}}
\newcommand*{\probacond}[2]{\mathop{\mathbb{P}}\left( #1 \, \mid \, #2 \right)}
\newcommand*{\esp}{\mathop{\mathbb{E}}}
\newcommand*{\var}{\mathop{\mathbb{V}}}
\newcommand*{\cov}{\mathop{\mathrm{cov}}}
\newcommand*{\corr}{\mathop{\rho}}

\newcommand*{\ber}{\mathop{\mathrm{Ber}}}
\newcommand*{\bin}{\mathop{\mathrm{Bin}}}
\newcommand*{\lnorm}{\mathop{\mathscr{N}}}
\newcommand*{\lexp}{\mathop{\mathcal{E}}}
\newcommand*{\lfisher}{\mathop{\mathscr{F}}}
\newcommand*{\lstu}{\mathop{\mathrm{t}}}
\newcommand*{\lgam}{\mathop{\Gamma}}
\newcommand*{\lunif}{\mathop{\mathscr{U}}}
\newcommand*{\lchisqr}{\mathop{\chi^2}}

\newcommand*{\sgn}{\mathop{\mathrm{sgn}}}
\newcommand*{\argmax}{\mathop{\mathrm{arg\,max}}}
\newcommand*{\argmin}{\mathop{\mathrm{arg\,min}}}
\newcommand*{\supp}{\mathop{\mathrm{supp}}}

	\thispagestyle{empty}
	\vspace*{-1.5cm}
	
	\begin{center}
        \vspace*{3cm}
		\vspace*{0.1cm}
		
		{\Huge On Constrained and $k$ Shortest Paths}
		\vspace*{1.5cm}
		
		{\large \textsc{Abderrahim Bendahi} \quad \quad \quad \quad \quad \textsc{Adrien Fradin}}
        
        {\footnotesize\texttt{\{adrien.fradin, abderrahim.bendahi\}@polytechnique.edu}}
		\vspace*{0.5cm}

        \textit{\monthyeardate\today}%

        \vspace*{0.1cm}
            
  
        \vspace*{2cm}
	\end{center}

\begin{adjustwidth}{3cm}{3cm}
\begin{abstract}
    Finding a shortest path in a graph is one of the most classic problems in algorithmic and graph theory. While we dispose of quite efficient algorithms for this ordinary problem (like the \textsc{Dijkstra} or \textsc{Bellman-Ford} algorithms), some slight variations in the problem statement can quickly lead to computati- \newline -onally hard problems. This article focuses specifically on two of these variants, namely the \textit{constrained shortest paths} problem and the $k$ \textit{shortest paths} problem. Both problems are NP-hard, and thus it's not sure we can conceive a polynomial time algorithm (unless $P = NP$), ours aren't for instance. Moreover, across this article, we provide ILP formulations of these problems in order to give a different point of view to the interested reader. Although we did not try to implement these on modern ILP solvers, it can be an interesting path to explore. 
    
    \hspace{3.5pt} We also mention how these algorithms constitute essential ingredients in some of the most important modern applications in the field of data science, such as \textit{Isomap}, whose main objective is the reduction of dimensionality of high-dimensional datasets.
\end{abstract}
\end{adjustwidth}
\vfill

	\section*{Notation}

		In the report, we adopt the following notations:
		\begin{itemize}
			\setlength\itemsep{0.1em}

			\item $G = (V, E, \omega)$: a directed, connected and weighted graph (with non-negative weights i.e. $\omega \colon E \to \R_+$). Here, graphs will be considered without loops or multiple edges. By convention, we note $n = \Card{V}$ and $m = \Card{m}$.

           \item $N^+(i)$ (resp. $N^-(i)$): the set of out-neighbors (resp. in-neighbors) of a vertex $i \in V$ in a directed graph $G = (V, E)$.
		\end{itemize}

  \clearpage%

  \section*{Introduction}

        \subsection{Motivations}%
            This project is about shortest paths algorithms and some variants of the original problem (two variants precisely, see parts \hyperlink{task2}{$2$} and \hyperlink{task3}{$3$}) in a weighted, directed and connected graph $G = (V, E)$ along with application, in the last part, to a dimensionality reduction strategy namely \textit{Isomap}.
        
            
            Finding shortest paths (in a network for instance) is a central problem in graph theory but also in computer science in general with wide applications in a broad range of fields (such as communication networks, logistics and even data science). Here, we will deal with some constrained shortest paths algorithms that are used in the \textit{Isomap} method, which aims at reducing the dimension of high-dimensional data sets while preserving the distance between each pair of data point (think of a data set composed of $100 \times 100$ pictures for instance, each sample have $10,000$ pixels and each pixel have $3$ channels; this result in a space whose dimension is $30,000$). Meanwhile reducing the dimension prevents some undesirable effects to arise with the curse of dimensionality, keeping the distances among data points is also essential if one wants to apply clustering methods (like $k d$-tree, NN graph...) for example.
            
            

            The project is split in four parts, the first three present different variants of the shortest paths problem while the fourth one is a comparison of the previous algorithms we implemented on task $1$ ($4$ algorithms\footnote{We did not use the parallelized version of the \textsc{Dijkstra}'s algorithm because its implementation is naive and irrelevant for our benchmark.}) and task $2$ ($2$ algorithms). Notice that only one algorithm is implemented in the third task so no comparison is possible\textellipsis

        \subsection{Overview}
            The project is split in several files and the code can be accessed on the \textsf{GitHub} page: \url{https://github.com/abderr03/On-Constrained-and-k-Shortest-Paths.git}. The main file is \texttt{main.cpp} which can be run (on a Unix environment) with the command:
            
                {\centering \texttt{./main data/input\_file.txt} \par}
                
            \noindent in the shell. It might be convenient to recompile the whole project using \texttt{make clean} followed by \texttt{make}. The code for the $i$-th task is in the file $i$\texttt{-task.cpp} while the file \texttt{tasks.hpp} contains all the macros and function headers. An \texttt{utils.cpp} file contains some useful functions used across all tasks. Finally, the folder \texttt{data} contains some of the files \texttt{rcsp}$X$\texttt{.txt} (where $X$ denotes a positive integer) along with some other hand-made graphs.

            The file used for the benchmark on task $1$ (resp. task $2$) is titled \texttt{benchmark\_1.py} (resp. \texttt{benchmark\_2.py}).

        \clearpage%
        \pagestyle{testpage}%
        \hypertarget{task1}{}%
        \section{The Single Source Shortest Paths problem (SSSP)}%
            We start by implementing two classical sequential algorithms for the SSSP problem (single-source-shortest-paths) on non-negatively weighted, directed and connected graphs which are \textsc{Dijkstra}'s algorithm (an example of a \textit{label-setting} algorithm) along with the \textsc{Bellman-Ford} algorithm (a \textit{label-correcting} algorithm). Note that \textsc{Dijkstra}'s algorithm only works in the non-negative weights setting, in which case it remains superior to the \textsc{Bellman-Ford} algorithms, even with improvements\footnote{See \href{https://www.ams.org/journals/qam/1970-27-04/S0033-569X-1970-0253822-7/}{Yen, Jin Y. (1970). \textit{An algorithm for finding shortest routes from all source nodes to a given destination in general networks}} or even \href{https://epubs.siam.org/doi/10.1137/1.9781611973020.6}{Bannister, M. J.; Eppstein, D. (2012). \textit{Randomized speedup of the Bellman–Ford algorithm}}.}. We also implemented a third algorithm which is the $\Delta$-stepping algorithm.
                

            Let $G = (V, E, \omega)$ be a graph satisfying all the previous discussed conditions and whose vertices are labeled from $1$ to $n$ that's say $V = \ens{1, 2, ..., n} = \Int1n$ (with the source vertex being noted as $s \in \Int1n$). We describe below the different sequential algorithms we implemented.

            Finally, notice that in the graph $G$, a shortest path from a source vertex $s$ to a target vertex $t$ with $s \neq t$ can always be taken as a \textit{simple path} i.e. the path does not go through the same vertex twice or more (otherwise the path contains a cycle $\mathscr{C}$ with non-negative weight which can just be removed). This remark also applies to the case when $s = t$ (a non-empty shortest path is then a cycle): except the endpoints of the cycle which are equal, we can always assume that neither the other crossed vertices appear twice or more nor they are equal to the endpoints. Actually, we can get rid of these \textit{non simple} shortest paths / cycles by requiring it to go through the minimum possible edges.
        
            \subsection{The \textsc{\textmd{Dijkstra}}'s algorithm}
                Below, the reader can find the rough pseudo-code of the \textsc{Dijkstra}'s algorithm. Actually, a key component of the algorithm does not explicitly appear, which is the data structure to use in order to find the right index $j$ (at line $8$) and update the distance array $d$ (see line $11$). A less critical part concerns the way we store the graph: throughout this project, the weighted graph $G$ will be stored as an \textbf{adjacency list} with pairs / tuples to keep track of some additional data held by the arcs (especially the weights, the delays -- for \hyperlink{task2}{task $2$} --...).

                The runtime complexity of the \textsc{Dijkstra}'s algorithm heavily depends on which data structure we use to keep track of the newest nearest vertex in the graph (see the above paragraph). In our case, we use a \textbf{priority queue}\footnote{Using \textbf{Fibonacci heap}, leads to the current best runtime complexity of $\O(m + n \log(n))$ where $n = \Card{V}$ and $m = \Card{E}$.} (the one provided in the \texttt{C++ STL} -- the standard library --), which supports the operations \textsc{FindMin} in time $\O(1)$, \textsc{DeleteMin} and \textsc{Insert} in time $\O(\log(n))$ where $n$ is the size of the priority queue. This priority queue does not provide any \textsc{DecreaseKey} method thus, our \textsc{Dijkstra}'s algorithm is implemented in a \textit{lazy fashion}\footnote{See: \href{https://nmamano.com/blog/dijkstra/dijkstra.html}{nmamano.com/blog/dijkstra/dijkstra.html} for some variants of the \textsc{Dijkstra}'s algorithm with its space and time complexity.} in which non-updated nodes still persist in the queue, at the cost of extra space usage.

                Our implementation have a theoretical $\O((m + n) \log(n))$ time complexity and a $\O(m)$ complexity in space. Also, we use an \textit{early break} approach by stopping the \textbf{while} loop right after we found the target vertex $t$ at the top of the heap.
                
						\begin{algorithm}
                                \DontPrintSemicolon
                                \SetKwProg{Init}{Initialization}{:}{}
                                \Init{}{
                                    $S \gets \ens{s}$ \tcp*[h]{The set of visited vertices.}\;
                                    $\textsf{d} \gets []$ \ \ \tcp*[h]{The array of the distance from $s$.}\;
                                    
                                }

                                \vspace*{\baselineskip}
                                \tcp*[h]{Pre-processing: fill the array $d$.}\;
                                \For{$i \in V$}{
                                    $\textsf{d}[i] = +\infty$\;
                                }

                                \vspace*{\baselineskip}
                                $\textsf{d}[s] = 0$\;
                                \While{$S \neq V$}{
                                    $j \gets \argmin\limits_{V \setminus S}(\textsf{d})$ i.e. $j \in V \setminus S$ and $\textsf{d}[j] = \min\limits_{i \in V \setminus S} \textsf{d}[i]$.\;

                                    \medskip

                                    $S \gets S \cup \left\{ j \right\}$\;
                                    \For{$k \in N^+(j)$}{
                                        \tcp*[h]{Relax the edge $(j, k)$ if necessary (if $k$ is already in $S$ this doesn't change the distance $\textsf{d}[k]$).}\;
                                        $\textsf{d}[k] \gets \min(\ens{\textsf{d}[k], \textsf{d}[j] + \omega(j, k)})$\;
                                    }
                                }
                            \caption{\textit{Sketch of the \textsc{Dijkstra}'s algorithm}}
							\label{alg:alg1}
                            \end{algorithm}

                Moreover, in order to obtain the vertices along a shortest path from $s$ to $t$, we maintain an array \textsf{pred} of the predecessors. It's enough to store only one predecessors per vertex as the \textsc{Dijkstra}'s algorithm aims at building the shortest path tree in $G$ rooted from $s$.

                In order to factorize our code, we put, in a separate \texttt{utils.cpp} file some redundant parts of code notably the \textsf{Relax} procedure (to relax an edge $(u, v) \in E$ and update $\textsf{d}$ and $\textsf{pred}$) and the \textsf{Path} procedure (to reconstruct a path given a vector of the predecessors). Note that this procedure along with the previous one have their own variant version for the constrained shortest path problem since the data structure are somehow different. These four procedures, fully implemented, can be found in the \texttt{utils.cpp} file.
                
                These pseudo-code of the \textsf{Relax} procedure is given here:

                \begin{algorithm}
                                \DontPrintSemicolon
                                \SetKwProg{Fn}{Function}{:}{}
			                         \SetKwFunction{FRelax}{Relax}

                        \Fn{\FRelax{$u$, $v$, $\textsf{d}$, $\textsf{pred}$}}{
                			\tcc{To relax the edge $(u, v)$ given the distance array $\textsf{d}$ and the array of predecessors $\textsf{pred}$.}
                				\If{$\textsf{d}[u] + \omega(u, v) < \textsf{d}[v]$}{
                                    $\textsf{d}[v] = \textsf{d}[u] + \omega(u, v)$\;
                					$\textsf{pred}[v] = u$\;
                                \KwRet \textbf{true}\;
                				}
                            \KwRet \textbf{false}\;
                			}

                   \caption{\textit{The \textsc{Relax} function}}
				    \label{alg:alg2}
                    \end{algorithm}

                    \clearpage%

                    Below the reader can find the pseudo-code of the \textsf{Path} procedure, this function returns \textbf{void} since it fill and modify in-place the (initially empty) vector \textsf{path}.
                    
                    \begin{algorithm}[H]
                                \DontPrintSemicolon
                                \SetKwProg{Fn}{Function}{:}{}
                            \SetKwFunction{FPath}{Path}

                        \Fn{\FPath{$s$, $t$, $\textsf{pred}$, $\textsf{path}$}}{
                			\tcc{To build in the array $\textsf{path}$ the path from $s$ to $t$ given the array of predecessors $\textsf{pred}$.}
                				\While{$s \neq t$}{
                                    $\textsf{path}.\textsc{Append}(t)$\;
                					$t \gets \textsf{pred}[t]$\;
                				}
                            $\textsf{path}.\textsc{Append}(s)$\;
                            $\textsc{Reverse}(\textsf{path})$\;
                			}
                                
                            \caption{\textit{The \textsc{Path}}}
							\label{alg:alg3}
                            \end{algorithm}  
                \noindent where $\textsc{Reverse}$ is a builtin function in \texttt{C++ STL} to reverse a vector. It would also be possible to use a \textbf{queue} or even a \textbf{deque} (a \textit{double-ended queue}) instead, that would prevent the \textsc{Reverse} operation at the end by performing \textsc{PushFront} operations in time $\O(1)$ but since \textsf{Path} is not a crucial function (contrary to the \textsf{Relax} procedure for instance) and its runtime is just $\O(n)$ i.e. linear in the number of vertices of the graph $G$, it can be neglected knowing other shortest path algorithms time complexity. The said optimizations are left to the interested reader.

                \subsection{The \textsc{\textmd{Bellman-Ford}} algorithm and some improvements}
                    Besides the well-known \textsc{Dijkstra}'s algorithm, we also implemented the \textsc{Bellman-Ford} algorithm. We have two version of this algorithm, a naive one, with a $\O(n m)$ runtime complexity and a more sophisticated one implementing some optimizations (found in the literature) which reduce the number of relaxations (they are well described at \cite{wikipedia-bellman-ford} and in \cite{bellman-ford-random}). The main idea of the \textsc{Bellman-Ford} algorithm is to relax repeatedly ($n - 1 = \Card{V} - 1$ times) all the edges $(u, v) \in E$ (intuitively, at each repetition of the main \textbf{for} loop, we \textit{propagate} the \textit{true} distance from the source in the graph). This algorithm is based on a dynamic programming approach, here is the pseudo-code:

                    \begin{algorithm}[H]
                                \DontPrintSemicolon
                                \SetKwProg{Init}{Initialization}{:}{}
                                \Init{}{
                                    $\textsf{d} \gets []$ \ \ \tcp*[h]{The array of the distance from $s$.}\;
                                    
                                }

                                \vspace*{\baselineskip}
                                \tcp*[h]{Pre-processing: fill the array $d$.}\;
                                \For{$i \in V$}{
                                    $\textsf{d}[i] = +\infty$\;
                                }

                                \medskip

                                $\textsf{d}[s] = 0$\;
                                \For{$i \in \Int{1}{\Card{V} - 1}$}{
                                    \For{$(u, v) \in E$}{
                                        \tcp*[h]{Relax the edge $(u, v)$ if necessary.}\;
                                        $\textsf{d}[v] \gets \min(\ens{\textsf{d}[v], \textsf{d}[u] + \omega(u, v)})$\;
                                    }
                                }
			                          
                            \caption{\textit{The \textsc{Bellman-Ford} algorithm}}
							\label{alg:alg2}
                            \end{algorithm}

                            The second version of the \textsc{Bellman-Ford} algorithm, named \texttt{bellman\_ford\_yen} in our project (in the name of Jin Y. Yen who mainly contributed to these improvements), rely on the following modifications:
                            \begin{itemize}
                                \item first, while some relaxations are still be performed, we continue to loop over the edges (this is done via a \textbf{do $\ldots$ while} loop) along with a boolean variable \textsf{relaxation}.
                                
                                \item on the other hand, we just need to relax edges $(u, v)$ when the distance $\textsf{d}[u]$ has been changed (in the current iteration or in the previous one) that's say, if $\textsf{d}[u]$ has not changed in some iteration, then there's no need to relax edges $((u, v))_{v \in N^+(u)}$ in the next iteration. We can track these vertices using two bool arrays: \texttt{to\_relax} (the vertices whose out-going edges need to be relaxed) and \texttt{queued} (the vertices $u \in V$ whose distance value $\textsf{d}[u]$ has changed in the current iteration).

                                \item finally, Jin Y. Yen noticed that it's better to partition the set of edges $E$ in two sets $E_1$ and $E_2$ such that:
                                    \[ E_1 := \enstq{(u, v) \in E}{u < v} \quad \text{(resp. $E_2 := \enstq{(u, v) \in E}{u > v}$ )} \]
                                and then to relax all out-going edges from $u = 1$, $\ldots$, $\Card{V}$ (in this order) which are in $E_1$ and then, do the same in reverse order for the edges in $E_2$ (now, $u = \Card{V}$, $\ldots$, $1$).
                            \end{itemize}
                        As observed by M. J. Bannister and D. Eppstein, we can substitute the \textit{natural} order $>$ on $V$ by a random permutation $\sigma \in \Sym(V)$ i.e. consider now:
                         \[ E_1 := \enstq{(u, v) \in E}{\sigma(u) < \sigma(v)} \quad \text{(resp. $E_2 := \enstq{(u, v) \in E}{\sigma(u) > \sigma(v)}$ )} \]
                        which reduce, on average, the number of iteration in the \textbf{do $\ldots$ while} loop. The interested reader can find in \cite{bellman-ford-random} the proofs along with pseudo-code of all the said optimization above.

                \subsection{The $\Delta$-stepping algorithm}
                    We have also implemented another sequential algorithm known as the $\Delta$-stepping which we don't know before but we find quite interesting to try implementing it. The algorithm is well describe in \cite{delta-stepping} with its pseudo-code (see page $123$, part $2$ -- \textit{The basic algorithm}). 

                     In our implementation we decided to use a \textbf{map} for the \textit{bucket priority queue}, where each non-empty buckets can be accessed through its index (a.k.a. priority) and its content is an \textbf{unordered set} (this prevents inserting multiple times the same vertex in a given bucket $\textsf{B}[i]$). Notice that in a map, in \texttt{C++}, the key are ordered in increasing order so that to access the bucket with minimum priority, it suffices to access to the \textit{first} bucket using the \texttt{begin()} method which returns an iterator pointing to the desired bucket.

                     \clearpage

                     For the sake of clarity, we describe below a slightly modified pseudo-code which better fits what we implemented in \texttt{C++}:

                    \begin{algorithm}[H]
                                \DontPrintSemicolon
                                \SetKwProg{Fn}{Function}{:}{}
			                         \SetKwFunction{FRelaxRequets}{RelaxRequests}

                            \SetKwProg{Init}{Initialization}{:}{}
                                \Init{}{
                                    $\textsf{d} \gets []$ \tcp*[h]{The array of the distance from $s$.}\;
                                    $E_{\ell} \gets \enstq{(u, v) \in E}{\omega(u, v) \leqslant \Delta}$ \tcp*[h]{The set of \textit{light} edges of $G$.}\;
                                    $E_{\mathcal{h}} \gets E \setminus E_{\ell}$ \tcp*[h]{The set of \textit{heavy} edges of $G$.}\;
                                    $\textsf{B} \gets []$ \tcp*[h]{The bucket priority queue.}\;
                                }

                            \vspace*{\baselineskip}

                        \Fn{\FRelaxRequets{$u$, $\Delta$, $E_0$, $\textsf{d}$, $\textsf{B}$}}{
                			\tcc{To relax the edge $(u, v)$ in $E_0$ and update the bucket priority queue \textsf{B}.}
                				\For(\tcp*[h]{$u$ is fixed.}){$(u, v) \in E_0$}{
                                    \tcp*[h]{Remove the vertex $v$ from its bucket.}\;
                                    \If{$\textsf{d}[v] < +\infty$}{
                                        $\textsf{B}\left[ \E{\frac{\textsf{d}[v]}{\Delta}} \right].\textsc{Remove}(v)$\;
                                    }
                                    \tcp*[h]{Relax the edge $(u, v)$ if necessary.}\;
                                    $\textsf{d}[v] \gets \min(\ens{\textsf{d}[v], \textsf{d}[u] + \omega(u, v)})$\;
                                        
                                    \tcp*[h]{Insert $v$ in the new bucket.}\;
                                    $\textsf{B}\left[ \E{\frac{\textsf{d}[v]}{\Delta}} \right].\textsc{Insert}(v)$\;
                                }
                			}

                        \medskip

                		$\textsf{B}[0].\textsc{Insert}(s)$\;
                        \While{$\textsf{B} \neq \varnothing$}{
                            $\textsf{R} \gets \varnothing$\;
                            $i \gets \textsf{B}.\textsc{FirstKey}()$\;
                            \Do{$\textsf{B}[i] \neq \varnothing$}{
                                $\textsf{B}_0 \gets \textsf{B}[i]$ \tcp*[h]{Perform a copy of the unordered set $\textsf{B}[i]$.}\;
                                $\textsf{R} \gets \textsf{R} \cup \textsf{B}[i]$\;
                                $\textsf{B}[i] \gets \varnothing$\;
                                 \For{$u \in \textsf{B}_0$}{
                                    $\textsf{RelaxRequests}(u, \Delta, E_{\ell}, \textsf{d}, \textsf{B})$\;
                                }
                            }

                            \medskip

                            \For{$u \in \textsf{R}$}{
                                $\textsf{RelaxRequests}(u, \Delta, E_{\mathcal{h}}, \textsf{d}, \textsf{B})$\;
                            }
                            $\textsf{B}.\textsc{DeleteKey}(i)$ \tcp*[h]{Remove the empty bucket of indes $i$.}\;
                        }
                                
                        \caption{\textit{The \textsc{$\Delta$-stepping} algorithm}}
						\label{alg:alg2}
                        \end{algorithm}

                    Notice that it's also possible to implement the bucket priority queue in various ways \cite{wikipedia-bucket-queue}. One approach that might worth a try is to replace the inner unordered sets by \textbf{queues} and maintain in a separate array the index of each vertex to prevent duplicate vertices being in the same bucket. In this way, one may try to prevent costly repeated copies of the unordered set $\textsf{B}[i]$ at line $17$ (but now, how can we still efficiently remove a given vertex from a bucket?). Another possibility one can also try is to use a \textbf{doubly-linked list} for the buckets and store, for each vertex, its right / left neighbors in the current bucket. We left to the interested reader these potential improvements.

                \clearpage%
                \hypertarget{task1-lp}{}%
                \subsection{A linear program viewpoint of the SSSP problem}
                    Interestingly, there exists an elegant linear program (LP) formulation of the SSSP problem, this formulation exists in many textbooks / articles in the literature (see \cite{b-doerr-inf421} or \cite{lp-network}). Given a directed, connected and weighted graph $G = (V, E, \omega$) with non-negative weights, the LP is as follow:
                    \begin{equation*}
                    \begin{array}{ll@{}ll}
                        \text{maximize}  & \displaystyle\sum\limits_{u \in V} d_u &\\
                        \text{subject to}& d_v - d_u \leqslant \omega(u, v) & \quad  \text{for all $(u, v) \in E$} \\
                                         & d_s = 0
                    \end{array}
                    \end{equation*}
                    where $(d_u)_{u \in V}$ are the variables and $s \in V$ is the source. Now, given an optimal solution $d^\star = (d_u^\star)_{u \in V}$ of the LP, one can recover the shortest path tree by using a \textit{back-tracking} approach like a Depth-First Search (DFS) in which we maintain the distance from the source $s$ to the current vertex and check which neighbor to choose according to the computed distances vector $d^\star$.

                    In the above LP, we do not explicitly know which edges have been chosen in the optimal solution and so, post-processing is needed to get the desired shortest paths. A more natural formulation of the SSSP leads to the following \textit{integer linear program} (ILP), which can be found for example in the book \cite{integer-programming}:
                    \begin{equation*}
                    \begin{array}{ll@{}ll}
                        \text{minimize}  & \displaystyle\sum\limits_{(u, v) \in E} x_{(u, v)} \omega(u, v) &\\
                        \text{subject to}& \displaystyle\sum_{u \in N^+(s)} x_{(s, u)} - \displaystyle\sum_{v \in N^-(s)} x_{(v, s)} = 1 & \\
                                         & \displaystyle\sum_{u \in N^+(t)} x_{(t, u)} - \displaystyle\sum_{v \in N^-(t)} x_{(v, t)} = -1 & \\
                                         & \displaystyle\sum_{v \in N^+(u)} x_{(u, v)} - \displaystyle\sum_{p \in N^-(u)} x_{(p, u)} = 0& \quad \text{for all $u \in V \setminus \ens{s, t}$}\\
                                         & x_{(u, v)} \in \ens{0, 1} & \quad \text{for all $(u, v) \in E$}
                    \end{array}
                    \end{equation*}
                    where $(x_{(u, v)})_{(u, v) \in E}$ are boolean variables (i.e. $0$ or $1$) indicating if we take the edge or not, $s$ (resp. $t$) is the source (resp. target) vertex with $s \neq t$. In the case $s = t$, if we want a non-empty shortest path from $s$ to $s$, we have two possibilities: 
                    \begin{itemize}
                        \item either reformulate the above ILP as follow:
                        \begin{equation*}
                        \begin{array}{ll@{}ll}
                            \text{minimize}  & \displaystyle\sum\limits_{(u, v) \in E} x_{(u, v)} \omega(u, v) &\\
                            \text{subject to}& \displaystyle\sum_{u \in N^+(s)} x_{(s, u)} = 1 & \\
                                             & \displaystyle\sum_{v \in N^+(u)} x_{(u, v)} - \displaystyle\sum_{p \in N^-(u)} x_{(p, u)} = 0& \quad \text{for all $u \in V$}\\
                                             & x_{(u, v)} \in \ens{0, 1} & \quad \text{for all $(u, v) \in E$}
                        \end{array}
                        \end{equation*}
                        which forces to take at least one edge whose origin is $s$. From the algorithmic viewpoint (see for instance the above pseudo-code) this can be achieved by adding an \textbf{if} statement testing whether or not the path from $s$ to $t$ has a positive weight before reaching the \textbf{return} statement.
     
                        \item or modify the input graph $G = (V, E)$ in the following way: since we just want a \textit{simple} shortest path, we add a new vertex $s' \notin V$ in $G$, we now remove all the edges $(v, s)$ with $v \in N^-(s)$ and insert the edges $(v, s')$ that's say the out-neighbors of $s$ now point to $s'$ and $N^-(s) = \varnothing$ in the new graph $G'$. For the sake of clarity, here is depicted the transformation:

\begin{figure}[H]
					\centering
					\begin{subfigure}{0.5\textwidth}
     \centering
							\begin{tikzpicture}
\begin{scope}[every node/.style={circle,draw, minimum size= 0.6cm, inner sep=0pt, outer sep=0pt}]
    \node (A) at (0,0) {$s$};
    \node (B) at (2,2) {$u_1$};
    \node (C) at (2,-2) {$u_p$};
    \node (D) at (-2,2) {$v_1$};
    \node (E) at (-2,-2) {$v_q$};
\end{scope}

\begin{scope}[>={Stealth[black]},
              every edge/.style={draw=red,very thick}]
    \path [->] (A) edge node {} (B);
    \path [->] (A) edge node {} (C);
    \path [->] (D) edge node {} (A);
    \path [->] (E) edge node {} (A);
    \path (B) -- node[auto=false]{\vdots} (C);
    \path (D) -- node[auto=false]{\vdots} (E);
\end{scope}
\end{tikzpicture}

						\end{subfigure}%
						\begin{subfigure}{0.5\textwidth}
							\centering
							\begin{tikzpicture}
\begin{scope}[every node/.style={circle,minimum size= 0.6cm, inner sep=0pt, outer sep=0pt,draw}]
    \node (A) at (0.5,0) {$s$};
    \node (A1) at (-0.5,0) {$s'$};
    \node (B) at (2,2) {$u_1$};
    \node (C) at (2,-2) {$u_p$};
    \node (D) at (-2,2) {$v_1$};
    \node (E) at (-2,-2) {$v_q$};
\end{scope}

\begin{scope}[>={Stealth[black]},
              every edge/.style={draw=red,very thick}]
    \path [->] (A) edge node {} (B);
    \path [->] (A) edge node {} (C);
    \path [->] (D) edge node {} (A1);
    \path [->] (E) edge node {} (A1);
    \path (B) -- node[auto=false]{\vdots} (C);
    \path (D) -- node[auto=false]{\vdots} (E);
\end{scope}
\end{tikzpicture}
						\end{subfigure}
      \caption{\textit{Input graph $G = (V, E)$ with a source vertex $s$ (left) and the new graph $G'$ with vertices $s$ and $s'$ (right)}.}
				\end{figure}
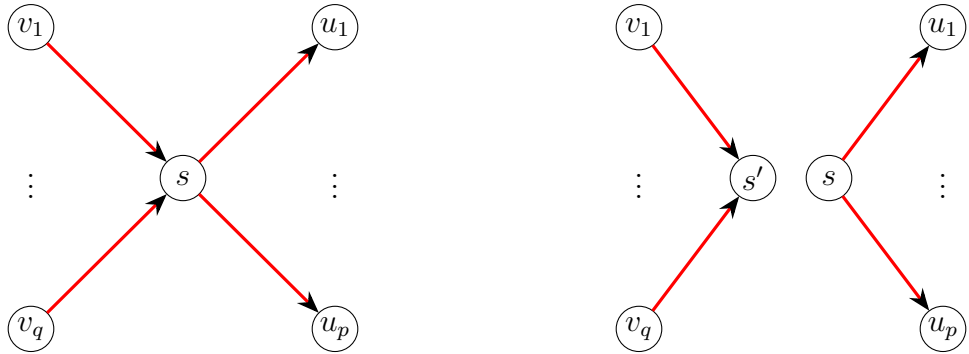
    thus, a non-empty shortest path in $G$ from $s$ to $s$ corresponds to a shortest path from $s$ to $s'$ in the new graph $G'$. This discussion will be helpful for the next task.
                    \end{itemize}
                    
        \clearpage%
        \hypertarget{task2}{}%
        \section{The Constraint Shortest Path problem}
            We slightly modify the above \textsc{Dijkstra} and \textsc{Bellman-Ford} algorithm by now taking care of a \textit{delay} constraints on each edge that is, we have two functions : 
                \[ \omega \colon E \to \R_+ \text{ and } d \colon E \to \R_+ \]
            and a bound $b \in \R_+$ (the first function being the weight (like in the previous part) and the second one being the \textit{delay} constraint). Our goal is to find a shortest path $\mathscr{C}$ from a source vertex $s$ to a target vertex $t$ such that $d(\mathscr{C}) \leqslant b$ i.e. the sum of the delays along the shortest path $\mathscr{C}$ does not exceed the bound $b$. Even in the case we have to find a constrained shortest path from $s$ to $s$, we just pre-process the graph $G$ as explained above (see paragraph \hyperlink{task1-lp}{$1.4$}) by adding a new vertex $s' \notin V$ so that we just have to find a constrained shortest path from $s$ to $s'$.
            
            Note that there's a fundamental difference between problem $1$ and problem $2$ (also, it appears that the problem $2$ is NP-hard \cite{np-hard} -- see problem \textbf{ND30} --) since, in a directed (connected) weighted graph $G$, if we know a shortest path from $s$ to $t \in V \setminus \ens{s}$ then, for each vertex $u$ along this path, we also know a shortest path from $s$ to $u$; this remark allows one to compute shortest paths step by step but, unfortunately, this does not necessarily hold true anymore when we add delay constraints on the edges. Consider for instance the following graph with source vertex $s$ and target vertex $s'$:
            \begin{figure}[H]
							\centering
							\begin{tikzpicture}
\begin{scope}[every node/.style={circle,minimum size= 0.6cm, inner sep=0pt, outer sep=0pt,draw}]
    \node (A) at (-1, 1) {$s$};
    \node (B) at (-1,-1) {$u$};
    \node (C) at ( 2, 0) {$v$};
    \node (D) at ( 5, 0) {$s'$};
\end{scope}

\begin{scope}[>={Stealth[black]},
              every edge/.style={draw=red,very thick}, every node/.style={color=blue, font=\bfseries}]
    \path [->] (A) edge[bend right = 30] node[left] {$(2, 1)$} (B);
    \path [->] (A) edge[bend left  = 20] node[above] {$(1, 5)$} (C);
    \path [->] (B) edge[bend right = 20] node[below] {$(1, 1)$} (C);
    \path [->] (C) edge node[above] {$(1, 1)$} (D);
\end{scope}
\end{tikzpicture}
      \caption{\textit{A counter-example to the above property, edges are labeled (weight, delay).}}
				\end{figure}
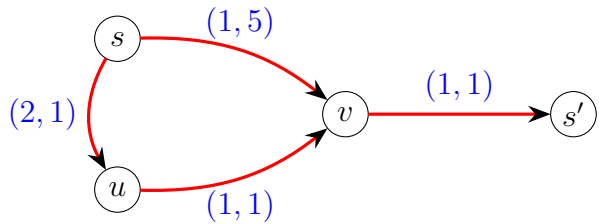
    Let's assume here that $b = 5$ and let the edge labels correspond to (weight, delay). A (constrained) shortest path from $s$ to $s'$ is the path $\mathscr{C} \colon s \rightarrow u \rightarrow v \rightarrow s'$ with weight $4 = 2 + 1 + 1$ and delay $3 = 1 + 1 + 1$ but, the sub-path $\mathscr{C}' \colon s \rightarrow u \rightarrow v$ is not a (constrained) shortest path from $s$ to $v$ anymore since its weight of $3$ exceeds the $1$-weighted path $\mathscr{C}'' \colon s \rightarrow v$. Hence, if we even try to compute all the shortest paths from $s$ to all other vertices, we won't necessarily obtain a tree anymore but rather a DAG (\textit{directed acyclic graph}) and finding shortest paths in such a DAG would require a DFS (\textit{depth-first search}) with a \textit{back-tracking} approach.

    Nonetheless, we have the following result: let $G = (V, E)$ be a directed (connected) and weighted graph equipped with a delay function $d \colon E \to \R_+$ (not to confuse with the vector / matrix of distances $\textsf{d}$ in the pseudo-code) and assume there exists a constrained shortest path $\mathscr{C}$ from a source vertex $s \in V$ to a target vertex $t \in V \setminus \ens{s}$. Let $u \in \mathscr{C}$ be any vertex along this path and denote $\mathscr{C}_u$ the truncated path $\mathscr{C}$ from $s$ to $u$ and fix:
        \[ \ell = d(\mathscr{C}_u) \]
    the delay of the truncated path, then, among all (simple) paths $\mathscr{P}$ in $G$ from $s$ to $u$ with $d(\mathscr{P}) \leqslant \ell$, $\mathscr{C}_u$ is a shortest path (otherwise, we could simply reject the truncated path and take another one minimizing the weight without violating the delay constraint). 
    
    This observation mimic in some way the property we observed for the SSSP problem and allows one to formulate a dynamic programming like recursion to compute the constrained shortest path distances. For this purpose, we introduce a matrix $\textsf{d}$ of size $n \times (b + 1)$ such that, given $u \in V$ and $\ell \in \Int0b$, $\textsf{d}[u][\ell]$ is the distance of a shortest path $\mathscr{P}$ (or $+\infty$ is none exists) from $s$ to $u$ with $d(\mathscr{P}) \leqslant \ell$. Here is the recursion satisfied by $\textsf{d}[\cdot][\cdot]$:
        \[ \textsf{d}[v][\ell] = \begin{cases} 0 & \text{if $ v = s$} \\ 
                                                \min\limits_{\substack{u \in N^-(v) \\ \ell \geqslant d(u, v)}} \left( \textsf{d}[u][\ell - d(u, v)] + \omega(u, v) \right) & \text{otherwise} \end{cases} \]
        for all $v \in V$ and $\ell \in \Int0b$. But because some delay might equal $0$, in order to get the correct distances in the matrix $\textsf{d}$, as in the \textsc{Bellman-Ford} algorithm, we have to apply the above recursion $\Card{V} - 1$ times on each column (starting from $\ell = 0$). Actually, we are more are less just repeatedly applying the \textsc{Bellman-Ford} algorithm to compute each column of $\textsf{d}$.
        
        One can thus implement a straightforward algorithm to compute the constrained shortest paths with runtime complexity $\O(n m b)$ (which is \textit{pseudo-polynomial} since it's exponential in the number of bits to represent the value of $b$):
        
        \begin{algorithm}[H]
                                \DontPrintSemicolon
                                \SetKwProg{Init}{Initialization}{:}{}
                                \Init{}{
                                    $\textsf{d} \gets [[]]$ \ \ \tcp*[h]{The $n \times (b + 1)$ matrix of the distances, with delays, from $s$.}\;
                                    
                                }

                                \vspace*{\baselineskip}
                                \tcp*[h]{Pre-processing: fill the matrix $d$.}\;
                                \For{$\ell \in \Int0b$}{
                                    \For{$u \in V$}{
                                        $\textsf{d}[u][\ell] \gets +\infty$\;
                                    }
                                    $\textsf{d}[s][\ell] = 0$\;
                                }

                                \medskip

                                \For{$\ell \in \Int0b$}{
                                    \For{$i \in \Int{1}{\Card{V} - 1}$}{
                                        \For{$(u, v) \in E$}{
                                            \If(\tcp*[h]{Ensure \textsf{delay} won't be negative.}){$d(u, v) \leqslant \ell$}{
                                                $\textsf{delay} \gets \ell - d(u, v)$\;
                                                \tcp*[h]{Relax the edge $(u, v)$ if necessary.}\;
                                                \If{$\textsf{d}[u][\ell] + \omega(u, v) < \textsf{d}[v][\textsf{delay}]$}{
                                                    $\textsf{d}[k][\textsf{delay}] \gets \textsf{d}[u][\ell] + \omega(u, v)$\;
                                            }
                                            }
                                        }
                                    }
                                }
			                          
                            \caption{\textit{An adaption of the \textsc{Bellman-Ford} algorithm for the constrained shortest path problem}}
							\label{alg:alg2}
                            \end{algorithm}

            \clearpage
            On the other hand, it's also possible to implement a \textsc{Dijkstra} variant to solve the constrained shortest path problem. We implemented the pseudo-code given below (be careful again not to confuse $\textsf{d}$, the distances matrix, with $d$ the delay function -- used at line $16$ --):
            
            \begin{algorithm}
                \caption{\textit{Sketch of the \textsc{Dijkstra}'s algorithm for the delay variant}}

                \DontPrintSemicolon
                                \SetKwProg{Init}{Initialization}{:}{}
                                \Init{}{
                                    \tcc{Recall $\textsf{d}[u][\ell]$ is the distance of a shortest path $\mathscr{P}$ (or $+\infty$ is none exists) from $s$ to $u$ with $d(\mathscr{P}) \leqslant \ell$.}
                                    $\textsf{d} \gets [[]]$ \tcp*[h]{The matrix $n \times (b + 1)$ of the distances from $s$.}\;
                                    $\mathscr{Q} \gets \varnothing$ \! \tcp*[h]{Priority queue of tuples $\ens{\text{delay}, \text{distance}, \text{vertex}}$.}\;
                                }

                                \vspace*{\baselineskip}
                                \tcp*[h]{Pre-processing: fill the array $d$.}\;
                                \For{$u \in V$}{
                                    \For{$\ell \in \Int0b$}{
                                        $\textsf{d}[u][\ell] \gets +\infty$\;
                                    }
                                }

                                \vspace*{\baselineskip}
                                $\textsf{d}[s][0] = 0$\;
                                $\mathscr{Q}.\textsc{Push}(\ens{0, 0, s})$\;
                                \While{$\mathscr{Q} \neq \varnothing$}{
                                    $\ens{\ell, \textsf{dist}, u} = \mathscr{Q}.\textsc{Top}()$\;
                                    $\mathscr{Q}.\textsc{Pop}()$\;

                                    \medskip

                                    \If{$u = s'$}{\KwRet\;}
                                    \If{$\textsf{dist} \leqslant \textsf{d}[u][\ell]$}{
                                    
                                    \For{$v \in N^+(u)$}{
                                        $\textsf{delay} \gets \ell + d(u, v)$\;
                                        \tcp*[h]{Relax the edge $(u, v)$ if necessary.}\;
                                        \If{$\textsf{delay} \leqslant b \ \wedge \ \textsf{d}[u][\ell] + \omega(u, v) < \textsf{d}[v][\textsf{delay}]$}{
                                                $\textsf{d}[k][\textsf{delay}] \gets \textsf{d}[u][\ell] + \omega(u, v)$\;
                                                $\mathscr{Q}.\textsc{Push}(\ens{\textsf{delay}, \textsf{d}[v][\textsf{delay}], v})$\;
                                            }
                                        }
                                    }
                                }
                                \tcp*[h]{No constrained shortest path from $s$ to $s'$ found...}\;
            \end{algorithm}

            Notice that, according to the previous ILP formulations (see part \hyperlink{task1}{$1$}), one can also formulate a (straightforward) ILP for this variant as follo:
            \begin{equation*}
                        \begin{array}{ll@{}ll}
                            \text{minimize}  & \displaystyle\sum\limits_{(u, v) \in E} x_{(u, v)} \omega(u, v) &\\
                            \text{subject to}& \displaystyle\sum_{u \in N^+(s)} x_{(s, u)} \geqslant 1 & \\
                                             & \displaystyle\sum_{v \in N^+(u)} x_{(u, v)} - \displaystyle\sum_{p \in N^-(u)} x_{(p, u)} = 0& \quad \text{for all $u \in V$}\\
                                             & \displaystyle\sum_{(u, v) \in E} x_{(u, v)} d(u, v) \leqslant b & \quad \text{the delay constraint}\\
                                             & x_{(u, v)} \in \ens{0, 1} & \quad \text{for all $(u, v) \in E$}\\
                        \end{array}
                        \end{equation*}
                         where $d \colon E \to \R_+$ is the delay function.

        \clearpage
        \hypertarget{task3}{}%
        \section{The $k$ Shortest Paths problem}

        Again, in this task we implement a variant of the \textsc{Dijkstra}'s algorithm to find the $k$-shortest paths from a given source vertex $s$. The key is to store all the paths that we have already visited while keeping in memory their total distance to the source (\textit{i.e.} the sum of weights of all the edges of the path). At each step, we select an unvisited path whose weight from the source is minimal, and then, we explore all the possible new paths by extending the current selected path using the out-neighbors and we update the total distances accordingly.
        
        The reader can find below the pseudo-code that we implemented:
        \begin{algorithm}
                \caption{\textit{Sketch of the \textsc{Dijkstra}'s algorithm for the $k$-shortest paths problem}}

                \DontPrintSemicolon
                                \SetKwProg{Init}{Initialization}{:}{}
                                \Init{}{
                                    $S \gets \ens{[0, s]}$ \tcp*[h]{The set of visited paths, each one being of the form [weight, $v_0$, $\ldots$, $v_n$].}\;

                                    $\mathscr{P} \gets []$ \quad \ \ \! \tcp*[h]{The array of the $k$-shortest paths.}
                                }

                                \vspace*{\baselineskip}
     
                                \While{$ \Card{P} < k$}{
                                    \tcp*[h]{With a priority queue for $S$, it can be done in $\O(1)$ time with \textsc{FindMin}.}\;
                                    Let $[d, s, \ldots, v] \in S$ such that $d = \min\limits_{\textsf{path} \in S} \textsf{path}[0]$\;

                                    \medskip
                                    
                                    $S \gets S \setminus \ens{[d, s, \ldots, v]}$\;
                                    \If(\tcp*[h]{To only append non-empty paths}){$\Card{[d, s, \ldots, v]} > 2$}{
                                        $\mathscr{P}.\textsc{append}([s, \ldots, v])$\;
                                    }

                                    \medskip

                                    \For{$k \in N^+(v)$}{
                                        
                                        $S.\textsc{append}([d + \omega(v, k), s, \ldots, v, k])$\;
                                    }
                                }
                                \Return $\mathscr{P}$ \;
                                
            \end{algorithm}

            Notice first that all the paths pushed in the set $S$ (implemented in our \texttt{C++} code as a priority queue) are distinct and at the end of the $i$-th iteration, all paths $\textsf{path} \in S$ satisfy $\Card{\textsf{path}} \leqslant i$\footnote{The size of a path is defined as the number of edges it contains.} (this can be proved by strong induction on the number of iterations on the \textbf{while} loop). For the first assertion for instance, if all paths in $S$ are distinct at some iteration $i$ then, let $\textsf{path}$ be the path at the top of the priority queue, at the next iteration, \textsf{path} is a prefix all the pushed paths and, if $S$ already contains a path of the form $\textsf{path} \cdot v$ -- where $v$ is a out-neighbor of the last vertex of $\textsf{path}$ -- then at some previous iteration $j < i$, the set $S$ must have contained $\textsf{path}$ two times or more, which is not possible by hypothesis -- this proves the heredity --, also, at the beginning $\Card{S} = 1$ so all path are distinct -- which gives the initialization step --.

            This bound on the size of the elements of $S$ at a given iteration $i$ show that the paths can't exceed a size of $k$: hence, the runtime complexity of this algorithm is $\O(n k^2)$ since $k$ iterations are performs in the \textbf{while} loop (line $4$) and at most $n$ iterations are performed in the \textbf{for} loop (line $9$) and in each of these iterations, we copy the current path $[d, s, \ldots, v]$, we modify it and we push it in $S$ which results in a cost of $\O(n k)$ thus, giving the $\O(n k^2)$ bound. Even if one try to improve the data structure and use, let say a bounded priority queue of size $k$ (or even a priority queue whose size decreases by one after each \textsc{Pop} operation), our current strategy of copying the paths won't give us a better worst-case runtime complexity than $\O(n k^2)$ in general (unless one assume properties on the input graph $G$ like a uniformly bounded out-degree by some constant $c < n$ for instance).

            On the other hand, we also try to find an ILP formulation of task $3$ but we didn't find one. Actually, the best we found so far is the following formulation:
            \begin{equation*}
                        \begin{array}{ll@{}ll}
                            \text{minimize}  & \displaystyle\sum_{i = 1}^k\sum\limits_{(u, v) \in E} x^{(i)}_{(u, v)} \omega(u, v) &\\
                            \text{subject to} & \displaystyle\sum_{u \in N^+(s)} x^{(i)}_{(s, u)} - \displaystyle\sum_{v \in N^-(s)} x^{(i)}_{(v, s)} = 1 - x^{(i)}_s & \quad \text{for all $i \in \Int1k$} \\
                                             & \displaystyle\sum_{v \in N^+(u)} x^{(i)}_{(u, v)} - \displaystyle\sum_{p \in N^-(u)} x^{(i)}_{(p, u)} = -x^{(i)}_u& \quad \text{for all $u \in V \setminus \ens{s}$ and all $i \in \Int1k$}\\
                                             & \displaystyle\sum_{(u, v) \in E} \abs{x^{(i)}_{(u, v)} - s^{(j)}_{(u, v)}} \geqslant 1 & \quad \text{for all $i \neq j$ in $\Int1k$} &\\
                                             & \displaystyle\sum_{u \in V} x^{(i)}_u = 1 & \quad \text{for all $i \in \Int1k$}\\
                                             & \displaystyle\sum_{u \in N^+(s)} x^{(i)}_{(s, u)} \geqslant 1 & \quad \text{for all $i \in \Int1k$}\\
                                             & x^{(i)}_{(u, v)} \in \N & \quad \text{for all $(u, v) \in E$ and all $i \in \Int1k$}\\
                        \end{array}
                        \end{equation*}
                         where $\left( x^{(i)}_{(u, v)} \right)_{\substack{(u, v) \in E \\ i \in \Int1k}}$ are non-negative integer variables indicating how many time we take the edge to build the $i$-th path and $\left(x^{(i)}_u \right)_{\substack{u \in V \\ i \in \Int1k}}$ are, with $s$, the other selected extremities of each of the $k$ paths respectively. Here, the critical point is the constraint:
                            \[ \sum_{(u, v) \in E} \abs{x^{(i)}_{(u, v)} - s^{(j)}_{(u, v)}} \geqslant 1 \]
                        whose purpose is to make sure each of the $k$ paths are distinct but it's not always true that two distinct paths always use a different set of edges (counting multiplicity here). This phenomenon arises when we begin to have cycles in the path (in which case, not only the set of edges is important but also the order in which we go through them). For example, consider the graph:
                         \begin{figure}[H]
							\centering
							\begin{tikzpicture}
\begin{scope}[every node/.style={circle,minimum size= 0.6cm, inner sep=0pt, outer sep=0pt,draw}]
    \node (A) at (-2, 0) {$u_0$};
    \node (B) at ( 0, 2) {$u_2$};
    \node (C) at ( 0, 0) {$u_1$};
    \node (D) at ( 0,-2) {$u_3$};
    \node (E) at ( 2, 0) {$u_4$};
\end{scope}

\begin{scope}[>={Stealth[black]},
              every edge/.style={draw=red,very thick}, every node/.style={color=blue, font=\bfseries}]
    \path [->] (A) edge node {} (C);
    \path [->] (C) edge node {} (E);
    \path [->] (C) edge[bend left  = 20] node {} (B);
    \path [->] (B) edge[bend left = 20] node {} (C);
    \path [->] (C) edge[bend right = 20] node {} (D);
    \path [->] (D) edge[bend right = 20] node {} (C);
\end{scope}
\end{tikzpicture}
      \caption{\textit{Two different paths with the same set of edges.}}
				\end{figure}
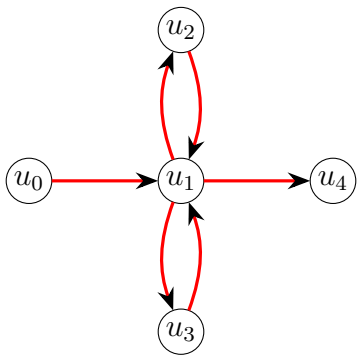

                        \noindent the two paths:
                            \[ \mathscr{C}_1 \colon u_0 \rightarrow u_1 \rightarrow u_2 \rightarrow u_1 \rightarrow u_3 \rightarrow u_1 \rightarrow u_4 \text{ and } \mathscr{C}_2 \colon u_0 \rightarrow u_1 \rightarrow u_3 \rightarrow u_1 \rightarrow u_2 \rightarrow u_1 \rightarrow u_4\] clearly use the same edges (the same number of times) but are distinct since they go through the two loops in different order. The above ILP doesn't care about the order in which we take the edges, it only guarantees that this set of edges correspond to one of the $k$ shortest path hence, $\mathscr{C}_1$ and $\mathscr{C}_2$ are considered \textit{equal} by the ILP.
                            
                            \begin{equation*}
                        \begin{array}{ll@{}ll}
                            \hspace*{-1.25cm}
                            \text{minimize}  & \displaystyle\sum_{j = 1}^k \left( \sum_{i = 1}^k\sum\limits_{(u, v) \in E} x^{(i, j)}_{(u, v)} \omega(u, v) \right) &\\
                           \hspace*{-1.25cm} \text{subject to}
											 & \displaystyle\sum_{(u, v) \in E} x^{(i, j)}_{(u, v)} \leqslant 1 && \text{for all $(i, j) \in \Int1k^2$} \\
											 & \displaystyle\sum_{(u, v) \in E} \left( x^{(i + 1, j)}_{(u, v)} - x^{(i, j)}_{(u, v)} \right) \geqslant 0 && \text{for all $j \in \Int1k$ and $i \in \Int{1}{k - 1}$} \\
											 & \displaystyle \sum_{v \in N^+(s)} x^{(1, j)}_{(s, v)} \geqslant 1 && \text{for all $j \in \Int1k$} \\
											 & \displaystyle \sum_{i = 1}^k \sum_{(u, v) \in E} \abs{x^{(i, j)}_{(u, v)} - x^{(i, \ell)}_{(u, v)}} \geqslant 1 && \text{for all $j \neq \ell$ in $\Int1k$} \\
											 & \displaystyle x^{(i, j)}_{(u, v)} \times \left( \sum_{\substack{(p, q) \in E \\ (p, q) \notin \enstq{(v, r) \in E}{r \in N^+(v)}}} x^{(i + 1, j)}_{(p, q)} \right) \leqslant 0 && \text{for all $(u, v) \in E$, $j \in \Int1k$ and $i \in \Int1{k - 1}$} \\
                                             & x^{(i, j)}_{(u, v)} \in \ens{0, 1} & & \text{for all $(u, v) \in E$ and all $(i, j) \in \Int1k^2$}\\
                        \end{array}
                        \end{equation*}

    \clearpage
    \section{Benchmarks}%
        \subsection{Evaluation methodology}%
            For each task, we choose to compare our algorithms on a single big graph which is described in the file \texttt{rcsp1.txt} (inside the \texttt{data/} folder). To ensure correctness when measuring the performance of each algorithm, we decided to select randomly $30$\footnote{We found this value of $30$ to be a good balance between the results we obtained and the time taken to run the whole benchmark.} distinct vertices
            \[\mathscr{V} = \ens{v_1, \ldots, v_{30}} \subset V = \Int{1}{100} \]
            then, for each algorithm $\mathscr{A}$ and for each pair $(v_i, v_j) \in \mathscr{V}^2$ with $i \neq j$ in $\Int{1}{30}$, we run $50$ times the selected algorithm $\mathscr{A}$ on the pair $(v_i, v_j)$. Over these runs, we measure $3$ \textit{times} (in nanoseconds):
            \begin{itemize}
                \item the total \textit{running time} (also called \textit{total time}) of the algorithm $\mathscr{A}$ (starting from the call to the function running algorithm $\mathscr{A}$)
                
                \item the \textit{pre-processing time}, the time performed by the algorithm $\mathscr{A}$ to do pre-computations and storage that will speed up the \textit{computation time}
                
                \item the \textit{computation time}, which is the time used by the algorithm $\mathscr{A}$ to end properly, after performing all the pre-computations
            \end{itemize}

            Then, we produce two kind of visualizations (a line plot and a heat-map) using each type of time (the \textit{pre-processing time}, the \textit{computation time} and the \textit{total time}):
            \begin{itemize}
                \item first, a line plot showing the average time taken by the $4$ algorithms over the weight of the shortest path found

                \item a heat-map (one for each algorithm), which is a fine-grained version of the above line plot, showing the running time (in $\mu \mathrm{s}$ -- microsecond --) evolution against the source and target vertices chosen in the \texttt{rcsp1.txt} graph
            \end{itemize}

        Some results are depicted below, for the heatmaps the grey cells refer to the case when the source vertex matches the target i.e. $i = j$ in our previous notation: we do not compute any time for that case.

        \subsection{Results for task $1$}%
       
        \begin{center}
            \includegraphics[scale=0.6]{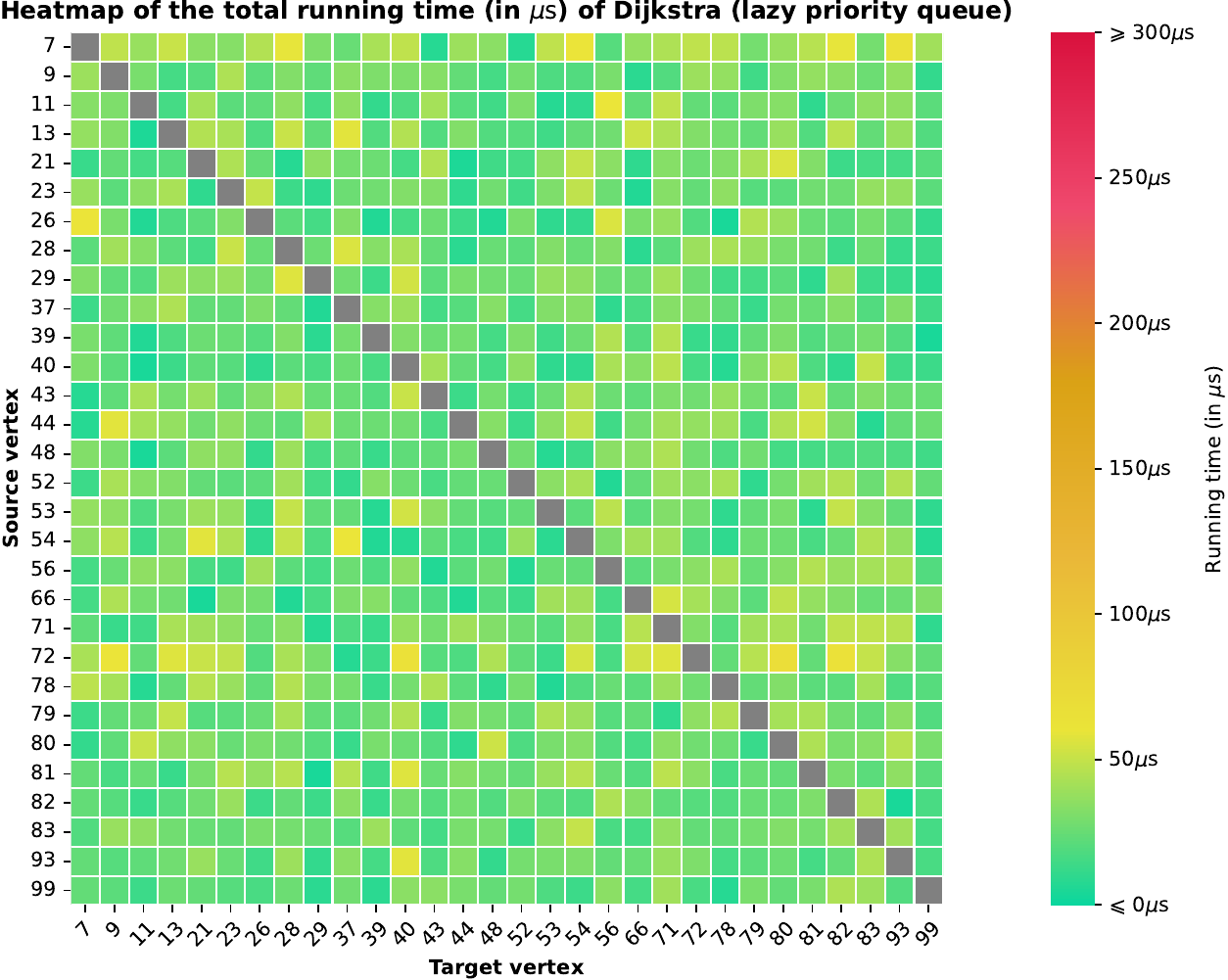}
            \captionof{figure}{\textit{Heatmap of the average total running time (in $\mathrm{\mu s}$) of \textsc{Dijkstra}'s algorithm}.}
        \end{center}

        \begin{center}
            \includegraphics[scale=0.6]{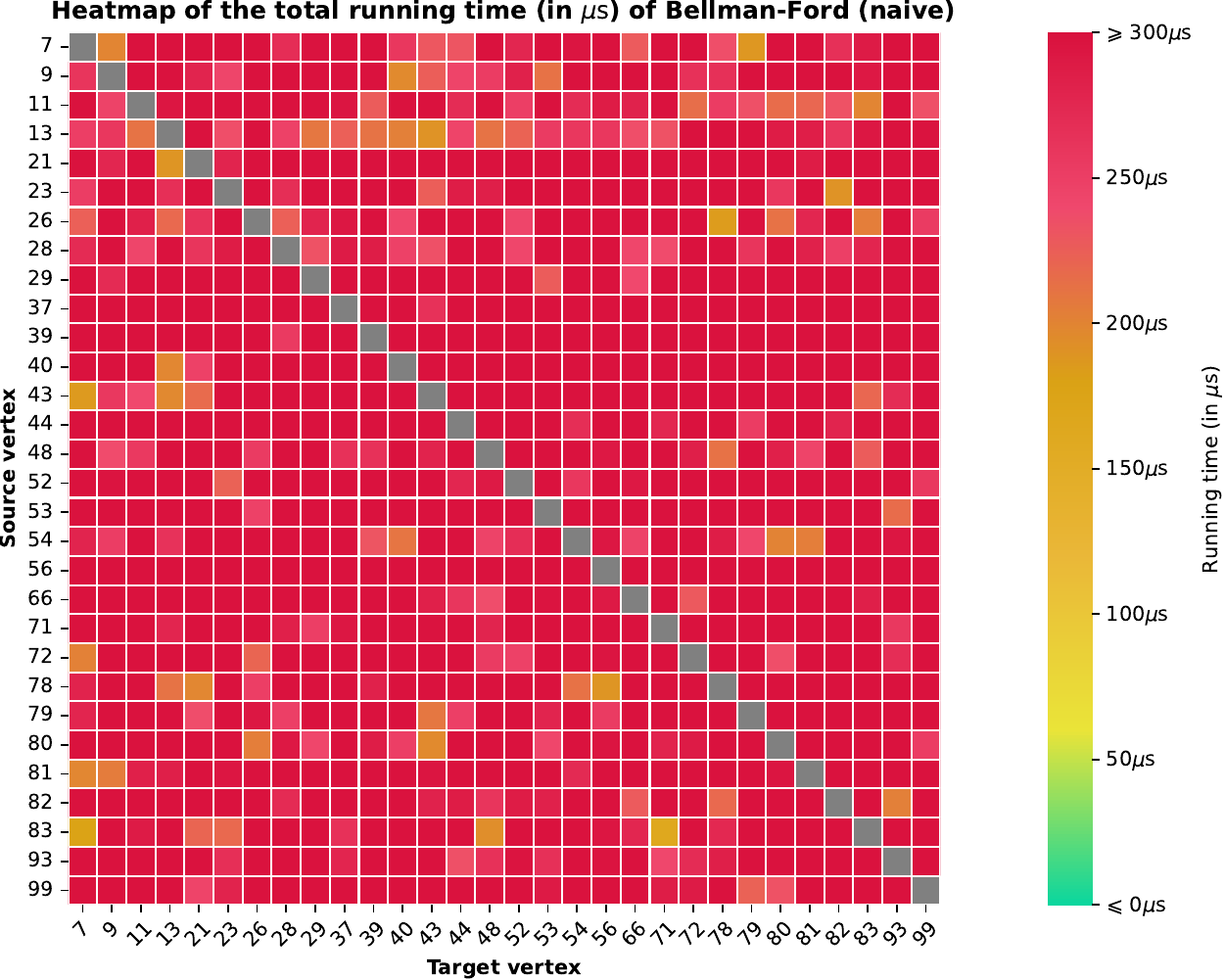}
            \captionof{figure}{\textit{Heatmap of the average total running time (in $\mathrm{\mu s}$) of naive \textsc{Bellman-Ford} algorithm}.}
        \end{center}
        
        \begin{center}
            \includegraphics[scale=0.6]{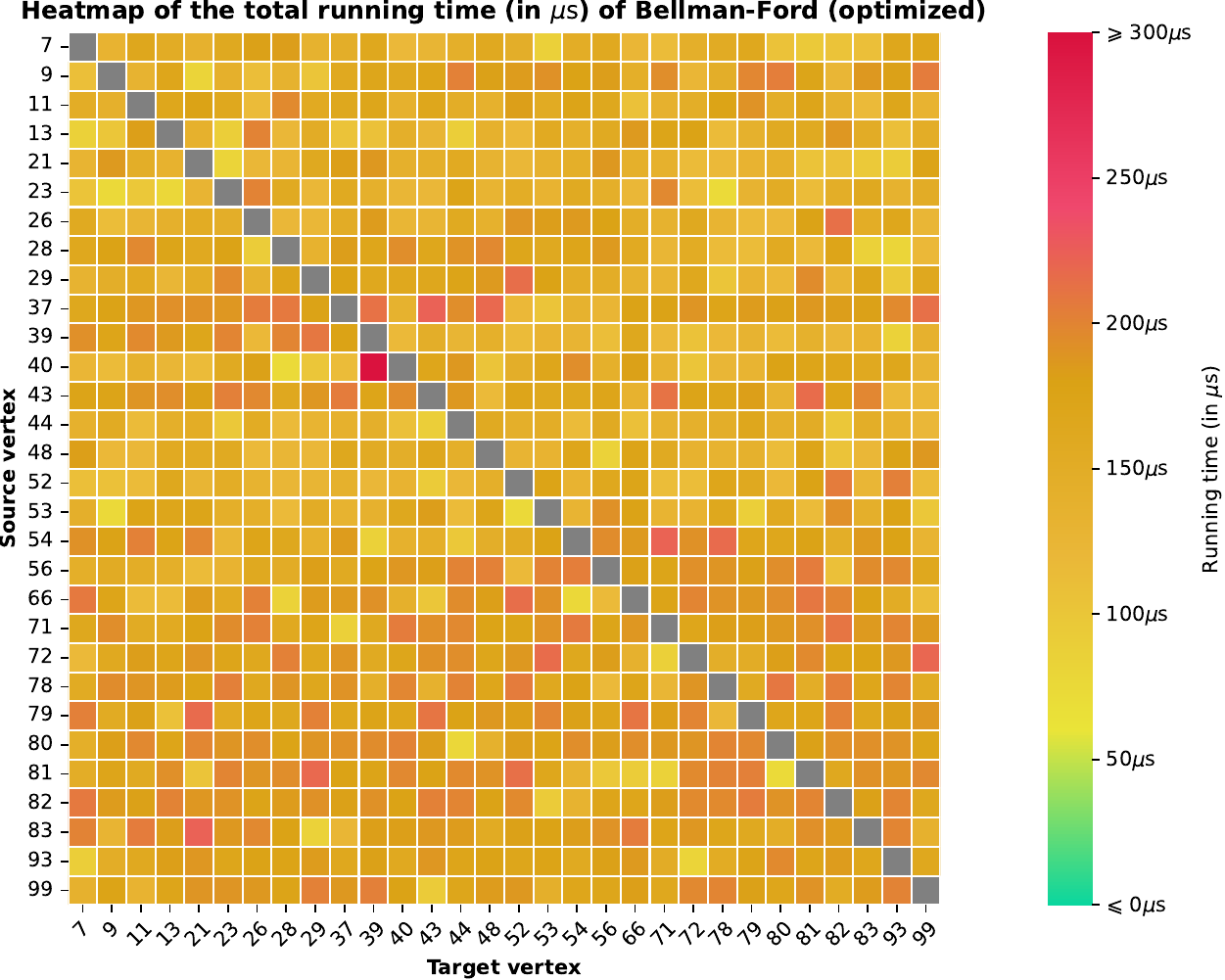}
            \captionof{figure}{\textit{Heatmap of the average total running time (in $\mathrm{\mu s}$) of optimized \textsc{Bellman-Ford} algorithm}.}
        \end{center}
        
        \begin{center}
            \includegraphics[scale=0.6]{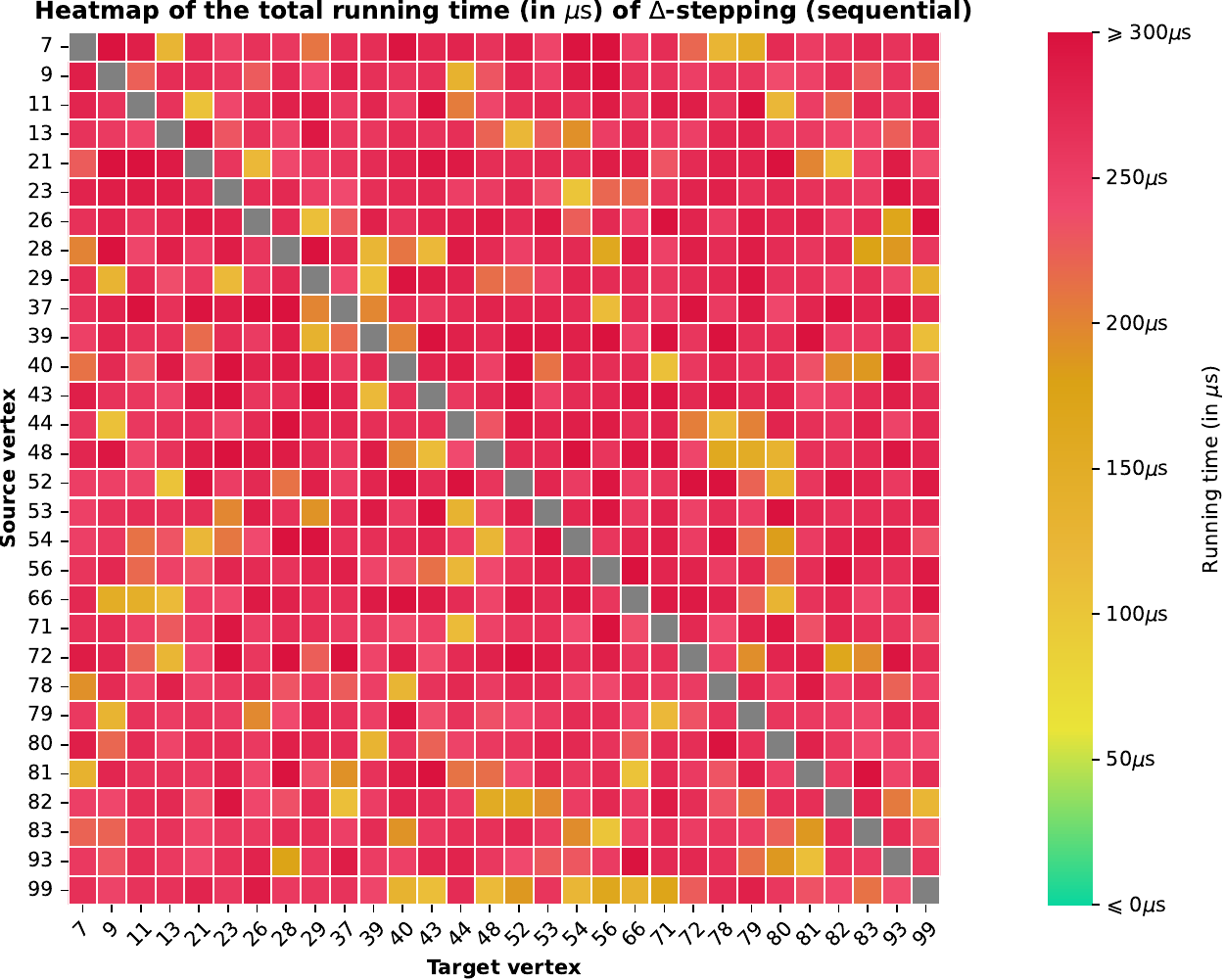}
            \captionof{figure}{\textit{Heatmap of the average total running time (in $\mathrm{\mu s}$) of \textsc{$\Delta$-stepping} algorithm}.}
        \end{center}

        \clearpage
        One can easily notice the following:
        \begin{itemize}
            \item while the \textsc{Dijkstra}'s algorithm demonstrates the best performance among all the implemented algorithms in terms of total running time, we notice that our optimized version of Bellman-Ford algorithm tends to have quite similar computation time performance for high values of distance (which we called \textit{path weight} in our plots)

            \item except \textsc{Dijkstra}'s algorithm, we observe that, surprisingly, the algorithm that has the best \textit{pre-processing time} (in this case the naive version of \textsc{Bellman-Ford}) has the worst computation time and total running time. This suggests that to be rewarded in computation, one must accept to make some sacrifices in \textit{pre-processing}
            
        \end{itemize}
        \subsection{Results for task $2$}%

        \begin{center}
            \includegraphics[scale=0.6]{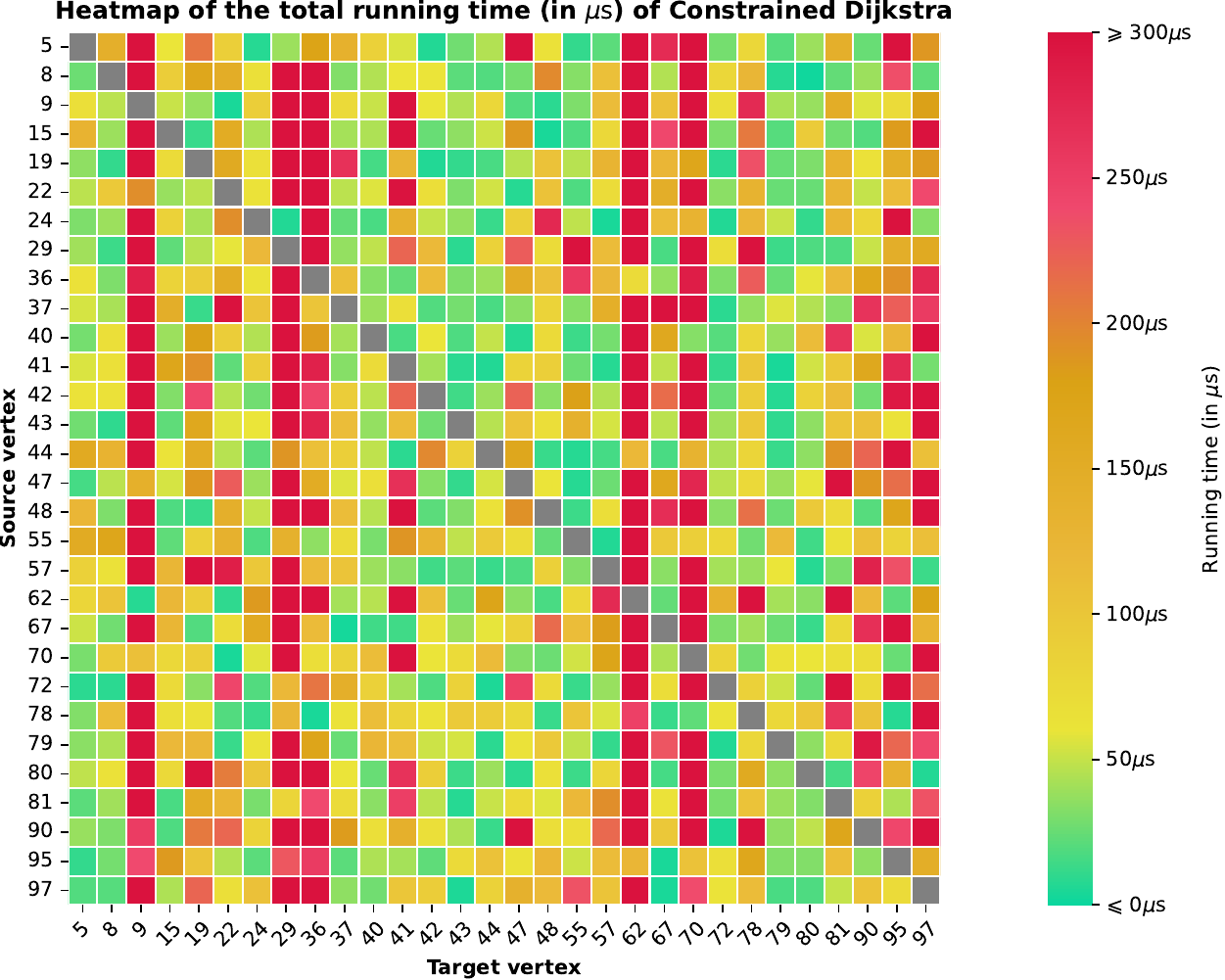}
            \captionof{figure}{\textit{Heatmap of the total running time (in $\mathrm{\mu s}$) of the constrained \textsc{Dijkstra}'s algorithm}.}
        \end{center}

        \begin{center}
            \includegraphics[scale=0.6]{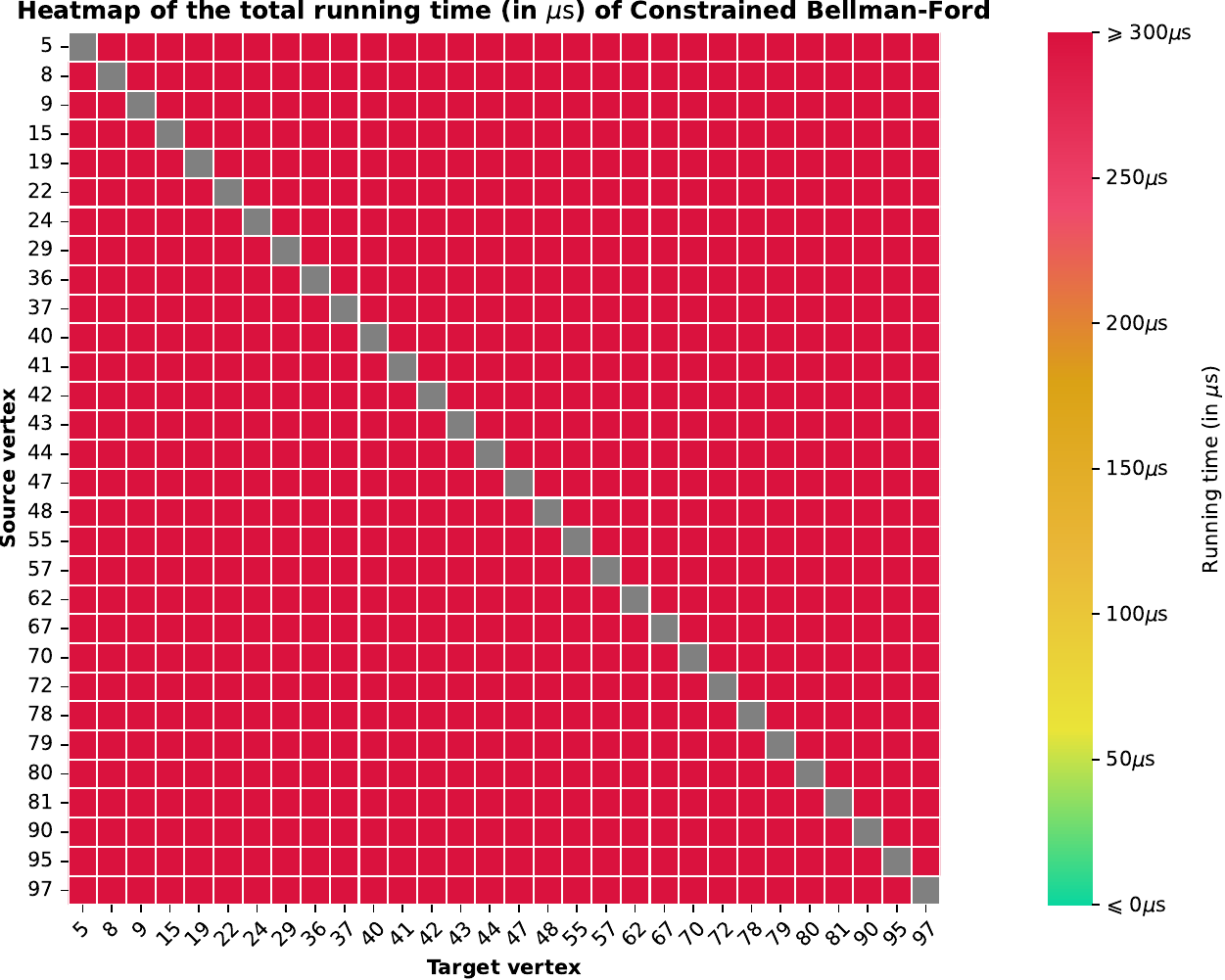}
            \captionof{figure}{\textit{Heatmap of the total running time (in $\mathrm{\mu s}$) of the constrained \textsc{Bellman-Ford}'s algorithm}.}
        \end{center}
        
        The same remarks apply for this second task. Except preprocessing time, \textsc{Dijkstra}'s constrained algorithm significantly outperforms the \textsc{Bellman-Ford} constrained algorithm.

\clearpage
\renewcommand{\refname}{References}%

	\phantomsection%
	\addcontentsline{toc}{section}{References}%

\clearpage%

\phantomsection%
\hypertarget{annex1}{}%
\section*{Annex A -- plots for task $1$}%

        \begin{center}
            \includegraphics[scale=0.6, angle = -90]{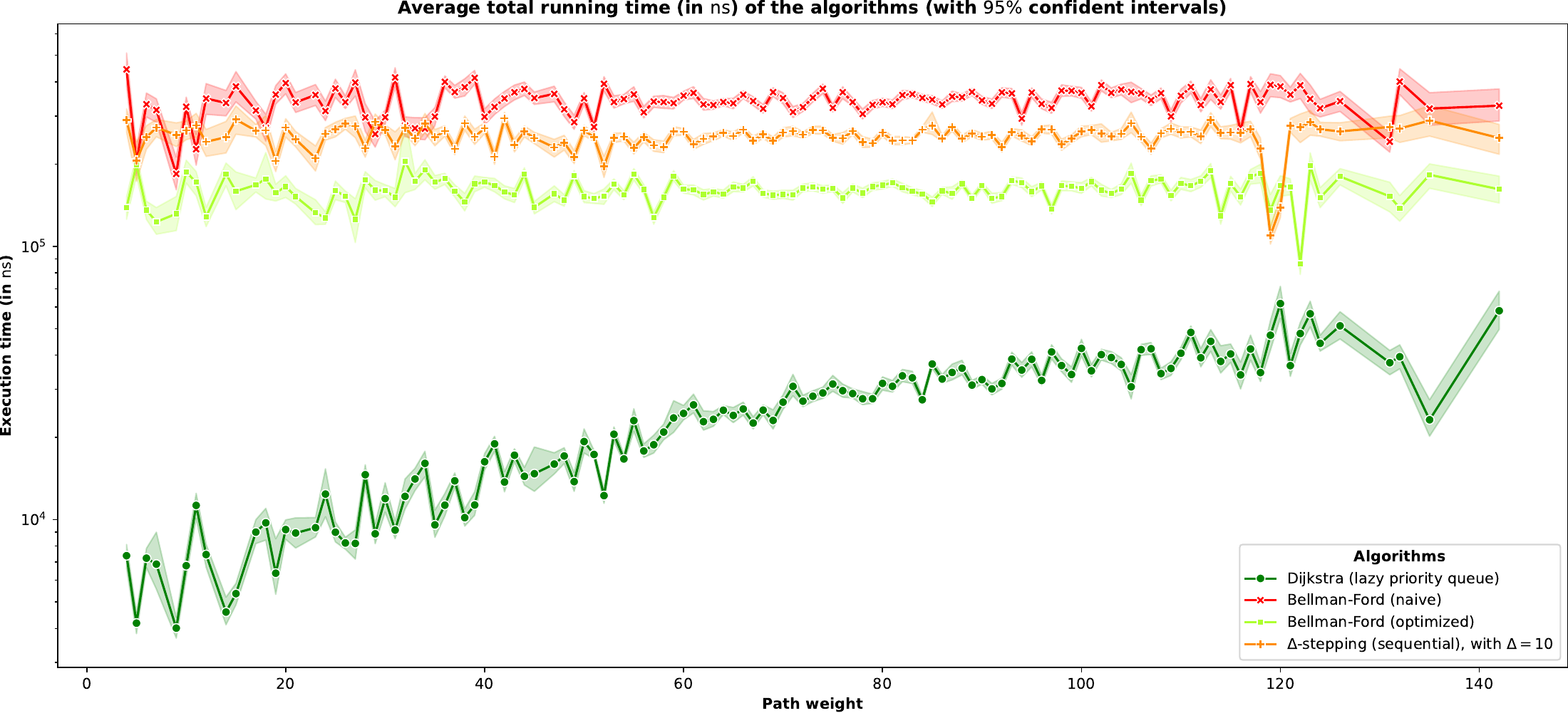}
            \captionof{figure}{\textit{Average total time (in $\mathrm{ns}$) of the four algorithms}.}
        \end{center}

        \begin{center}
            \includegraphics[scale=0.6, angle = -90]{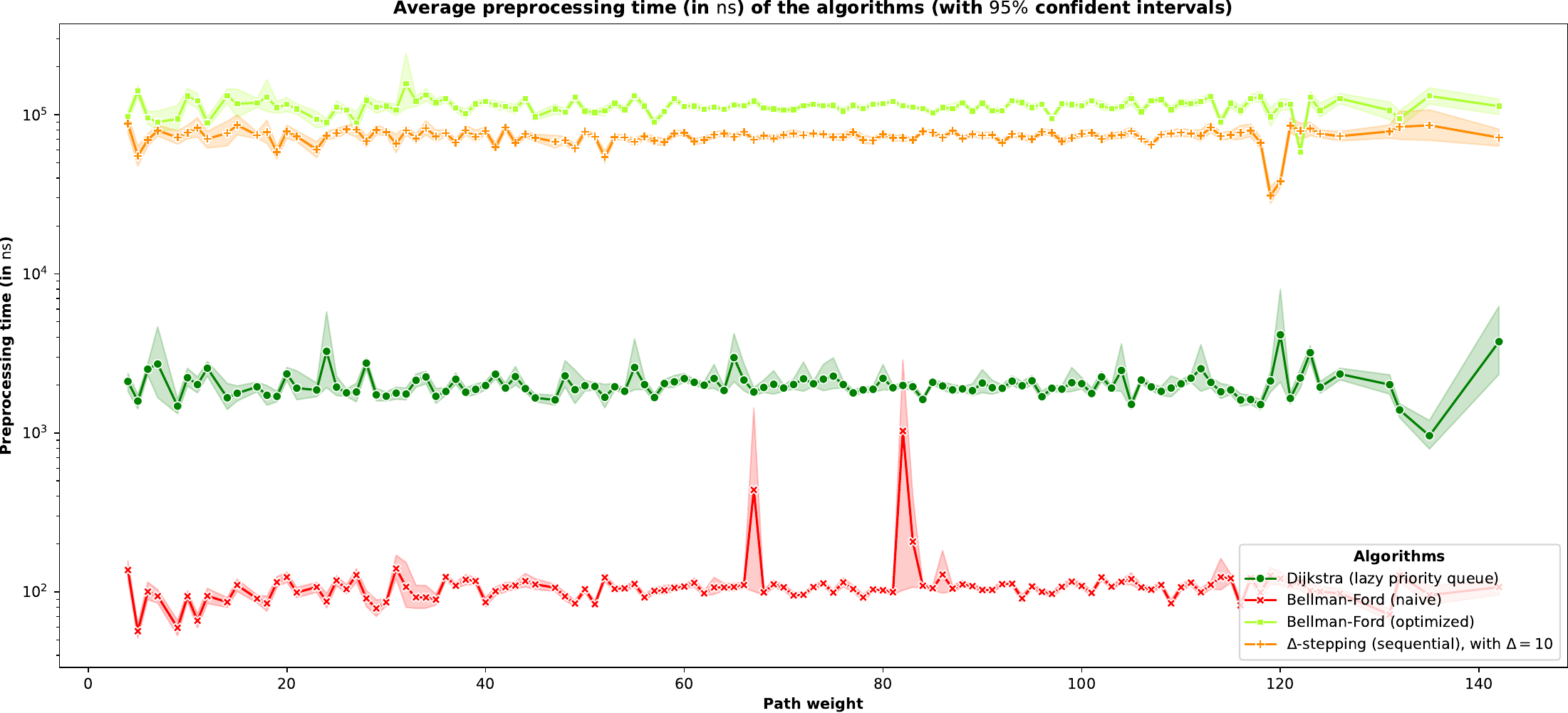}
            \captionof{figure}{\textit{Average pre-processing time (in $\mathrm{ns}$) of the four algorithms}.}
        \end{center}

        \begin{center}
            \includegraphics[scale=0.6, angle = -90]{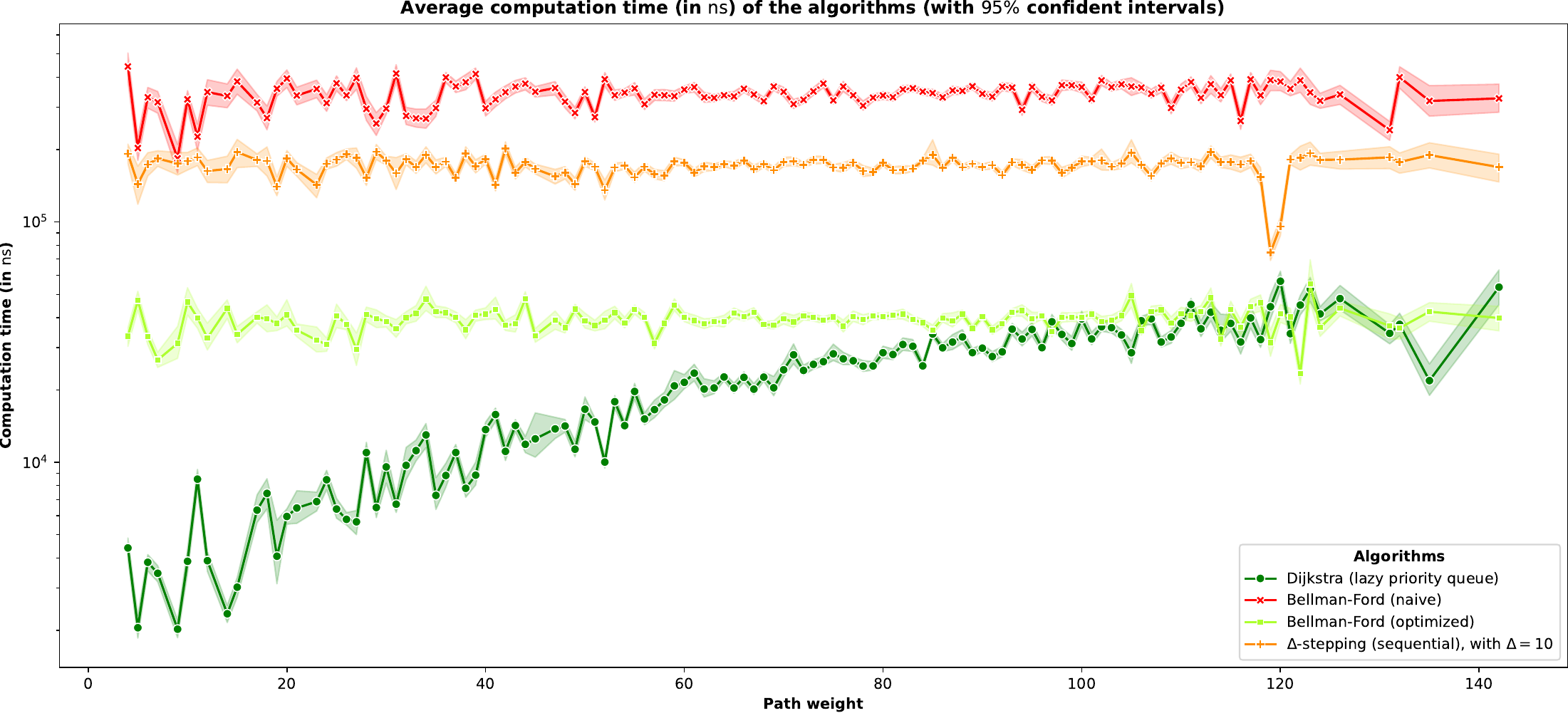}
            \captionof{figure}{\textit{Average computation time (in $\mathrm{ns}$) of the four algorithms}.}
        \end{center}

         \begin{center}
         \begin{figure}[H]
            \includegraphics[scale=0.65]{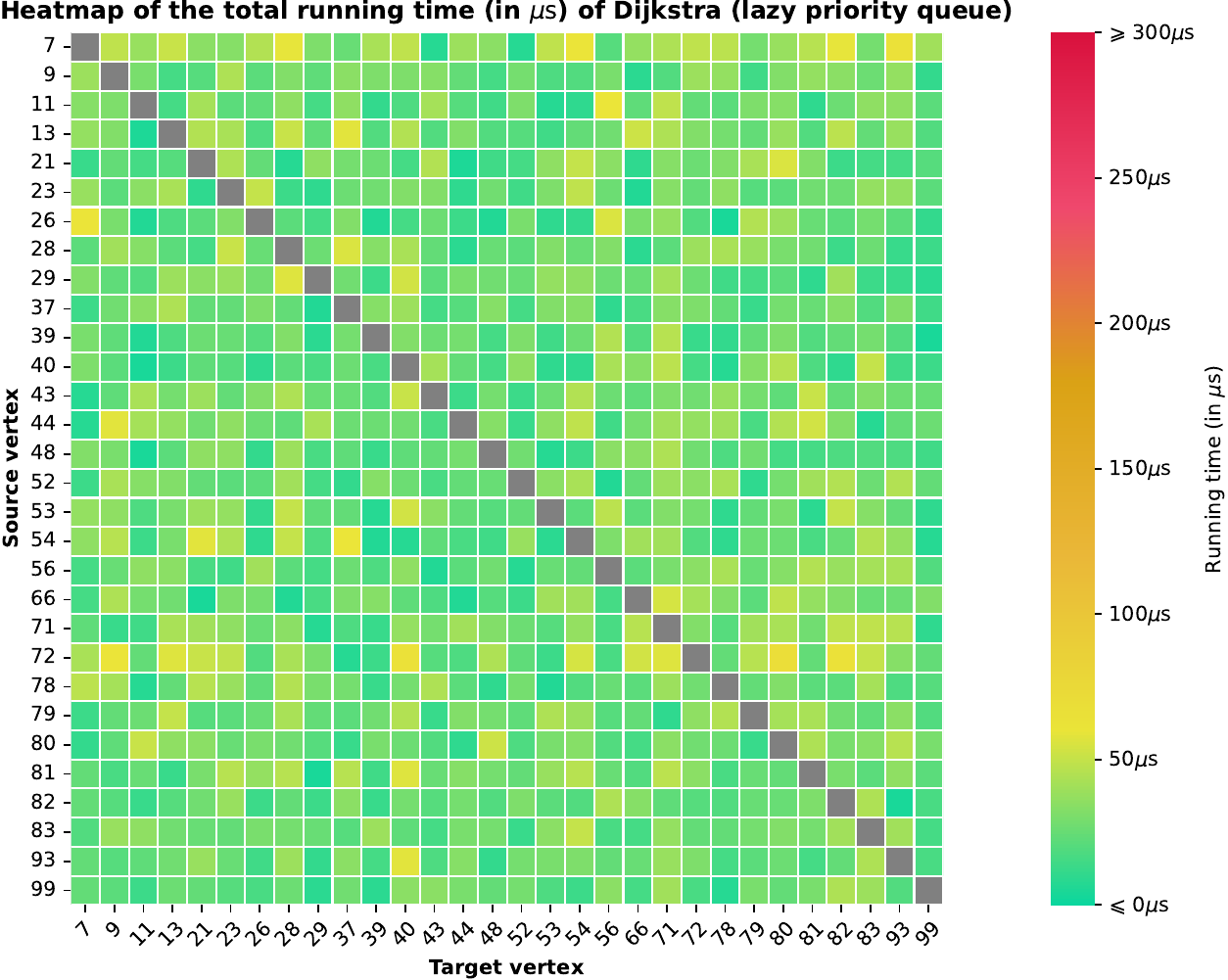}
            
            \vspace{0.3cm}
            
            \includegraphics[scale=0.65]{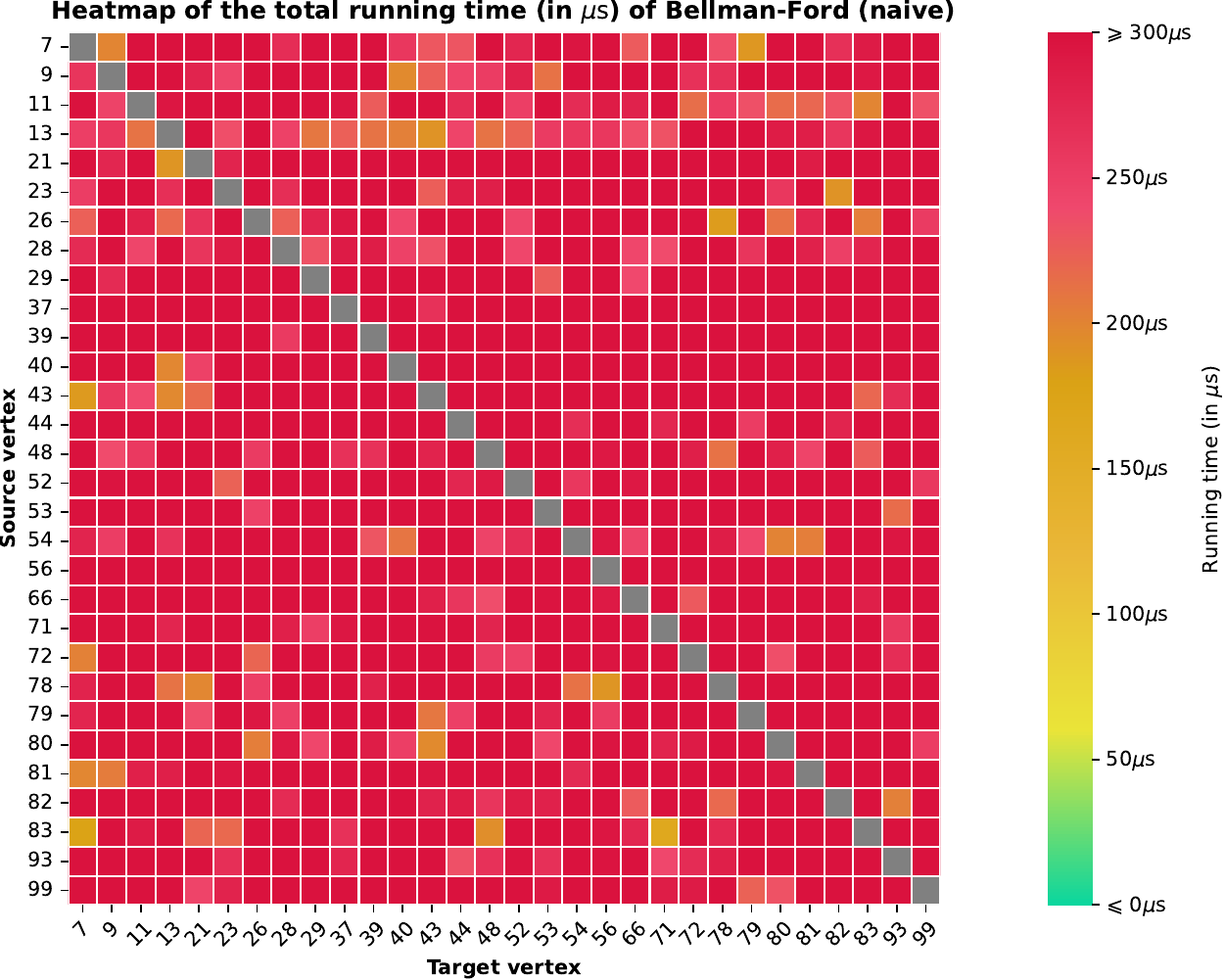}
            \captionof{figure}{\textit{Heatmaps of the total running time for \textsc{Dijkstra}'s and naive \textsc{Bellman-Ford} algorithm}.}
            \end{figure}
        \end{center}
         \begin{center}
         \begin{figure}[H]
            \includegraphics[scale=0.65]{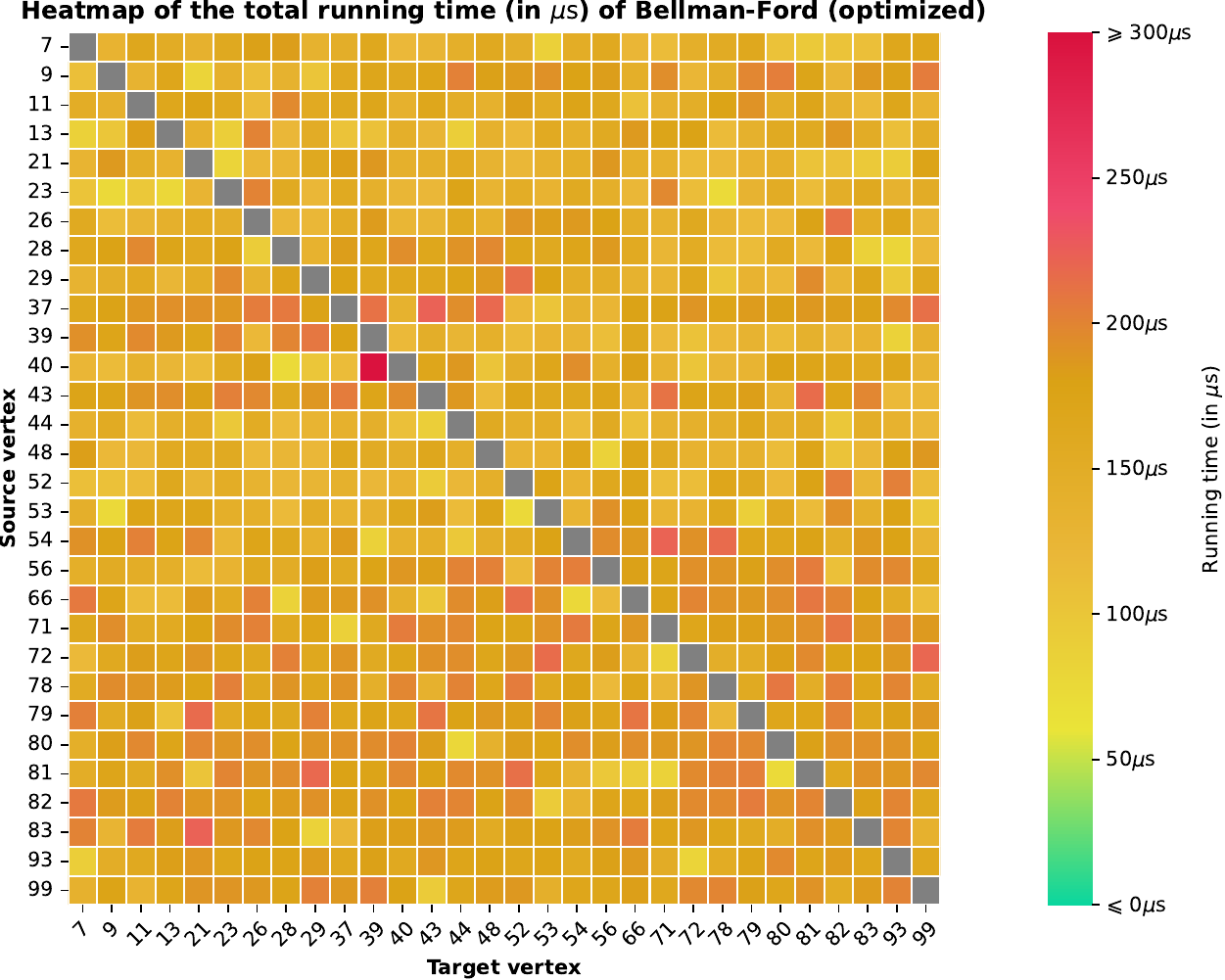}
            
            \vspace{0.3cm}
            
            \includegraphics[scale=0.65]{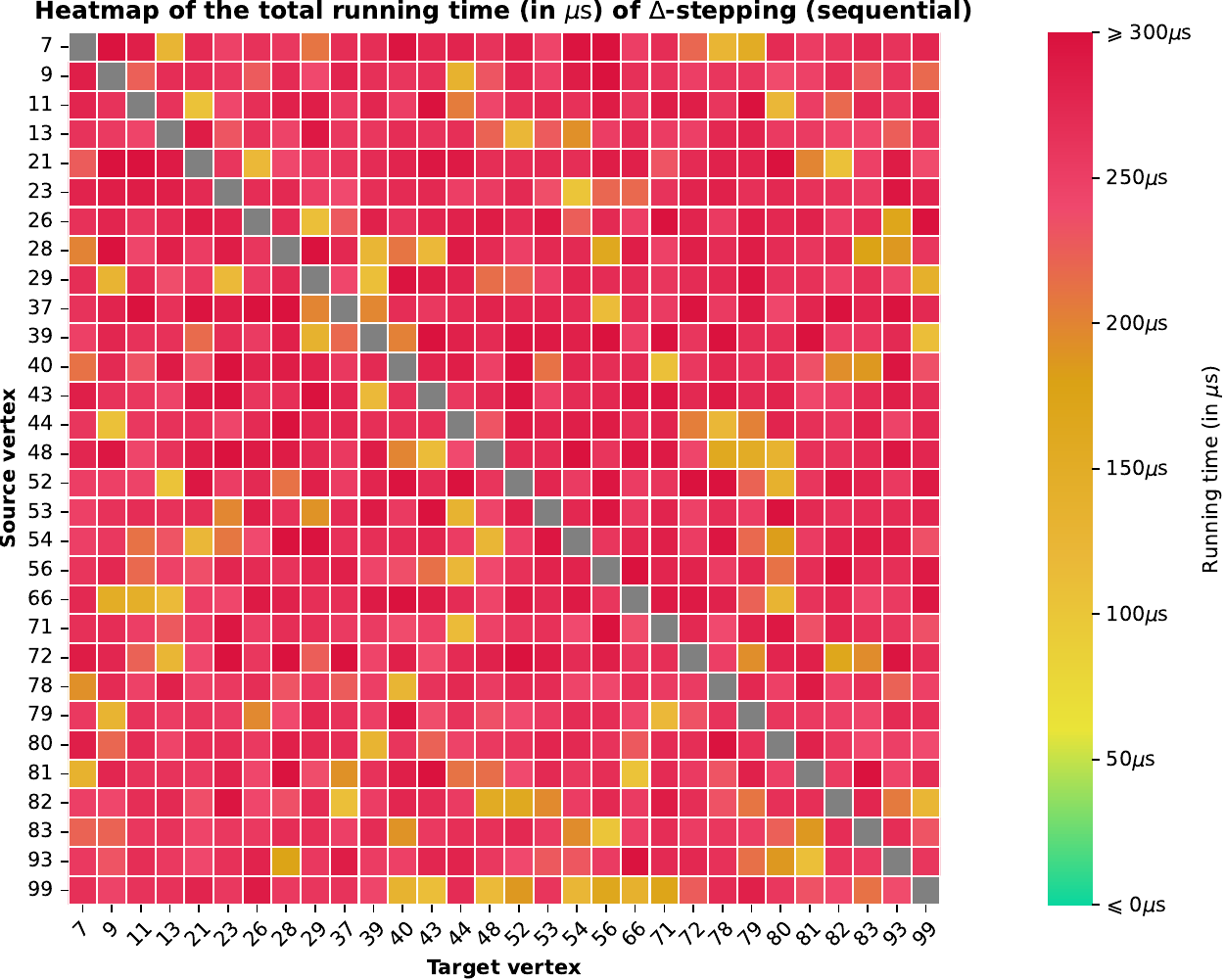}
            \captionof{figure}{\textit{Heatmaps of the total running time for optimized \textsc{Bellman-Ford}'s and \textsc{$\Delta$-stepping} algorithm}.}
            \end{figure}
        \end{center}

       \begin{center}
         \begin{figure}[H]
            \includegraphics[scale=0.65]{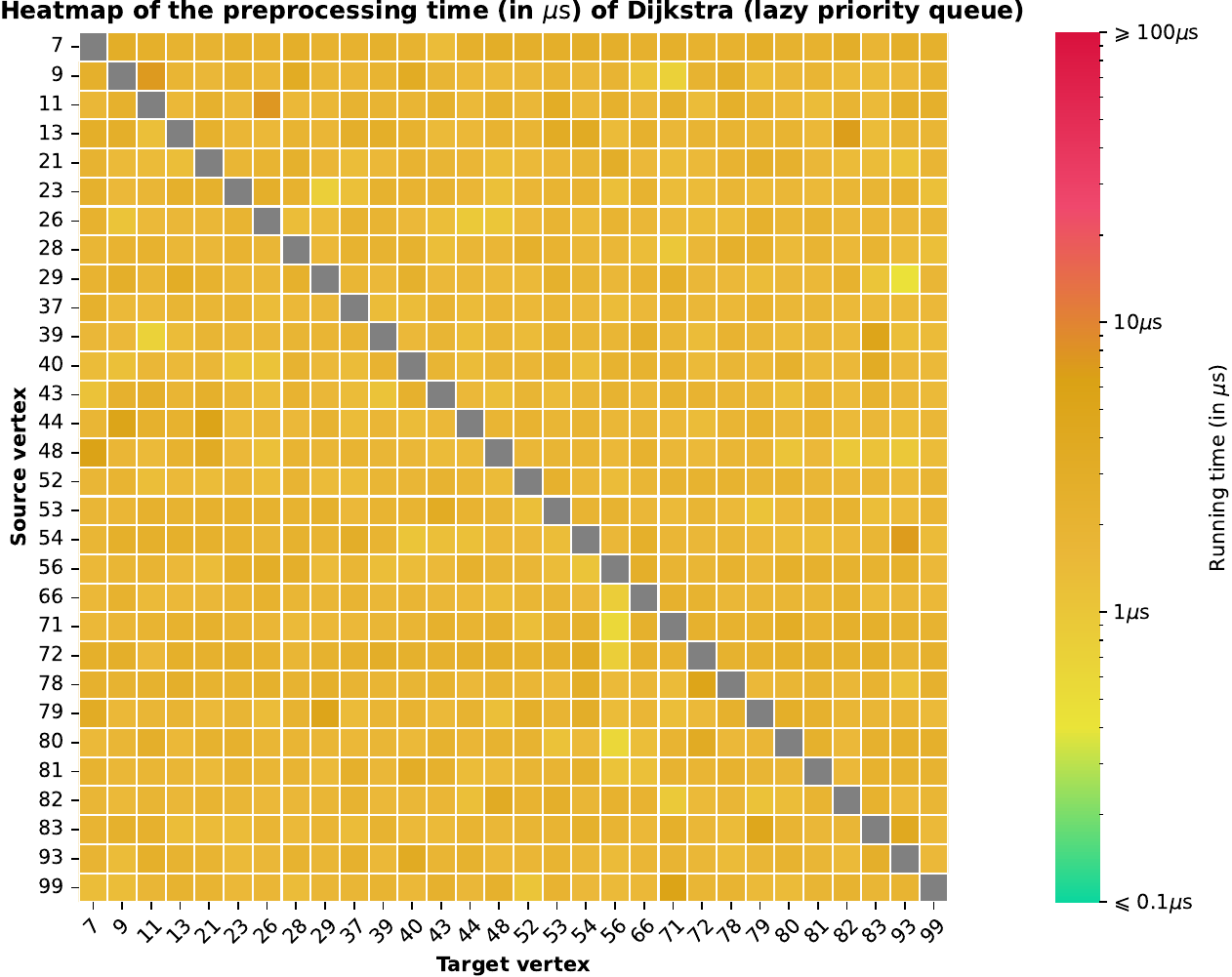}
            
            \vspace{0.3cm}
            
            \includegraphics[scale=0.65]{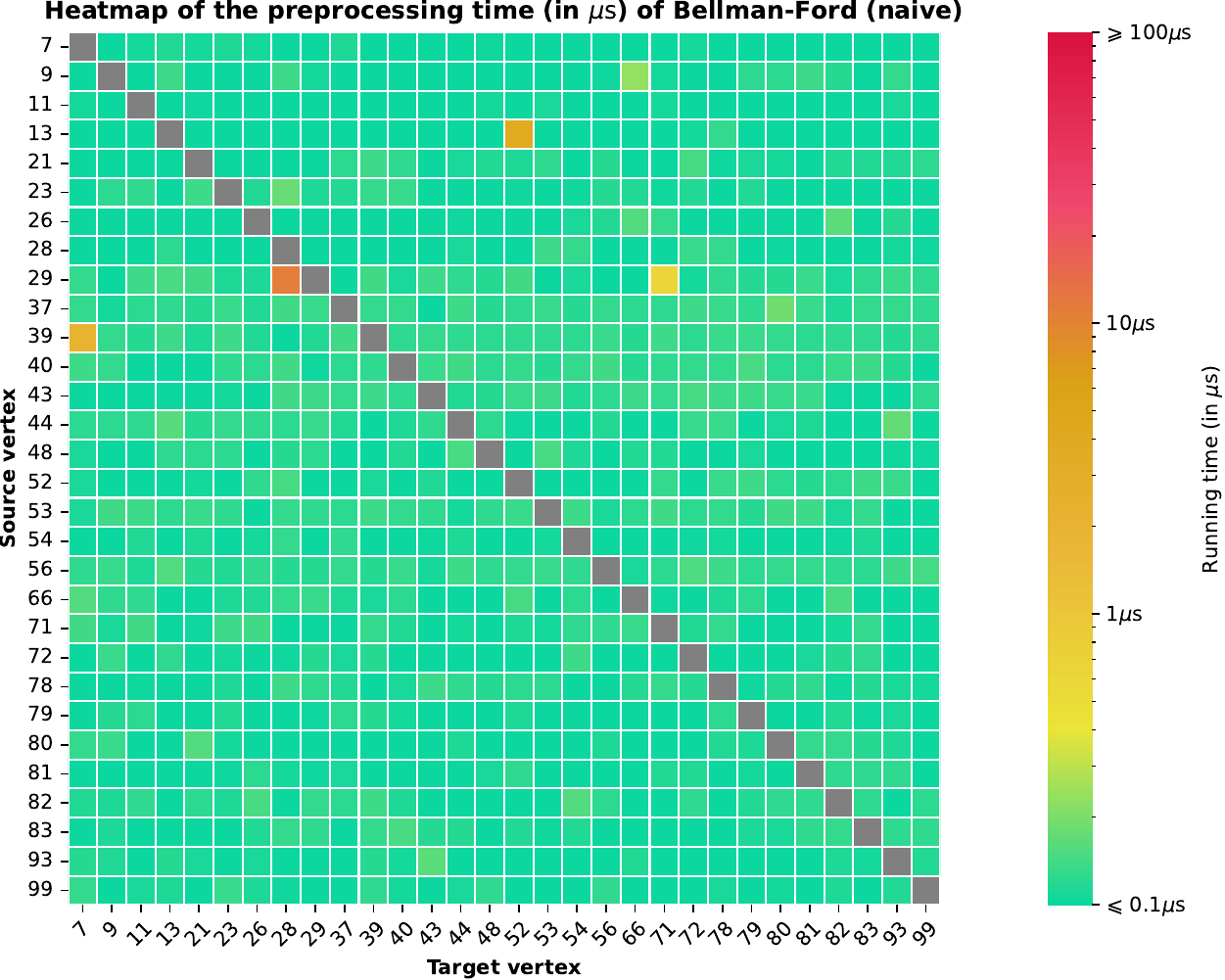}
            \captionof{figure}{\textit{Heatmaps of the pre-processing time for \textsc{Dijkstra}'s and naive \textsc{Bellman-Ford} algorithm}.}
            \end{figure}
        \end{center}

         \begin{center}
         \begin{figure}[H]
            \includegraphics[scale=0.65]{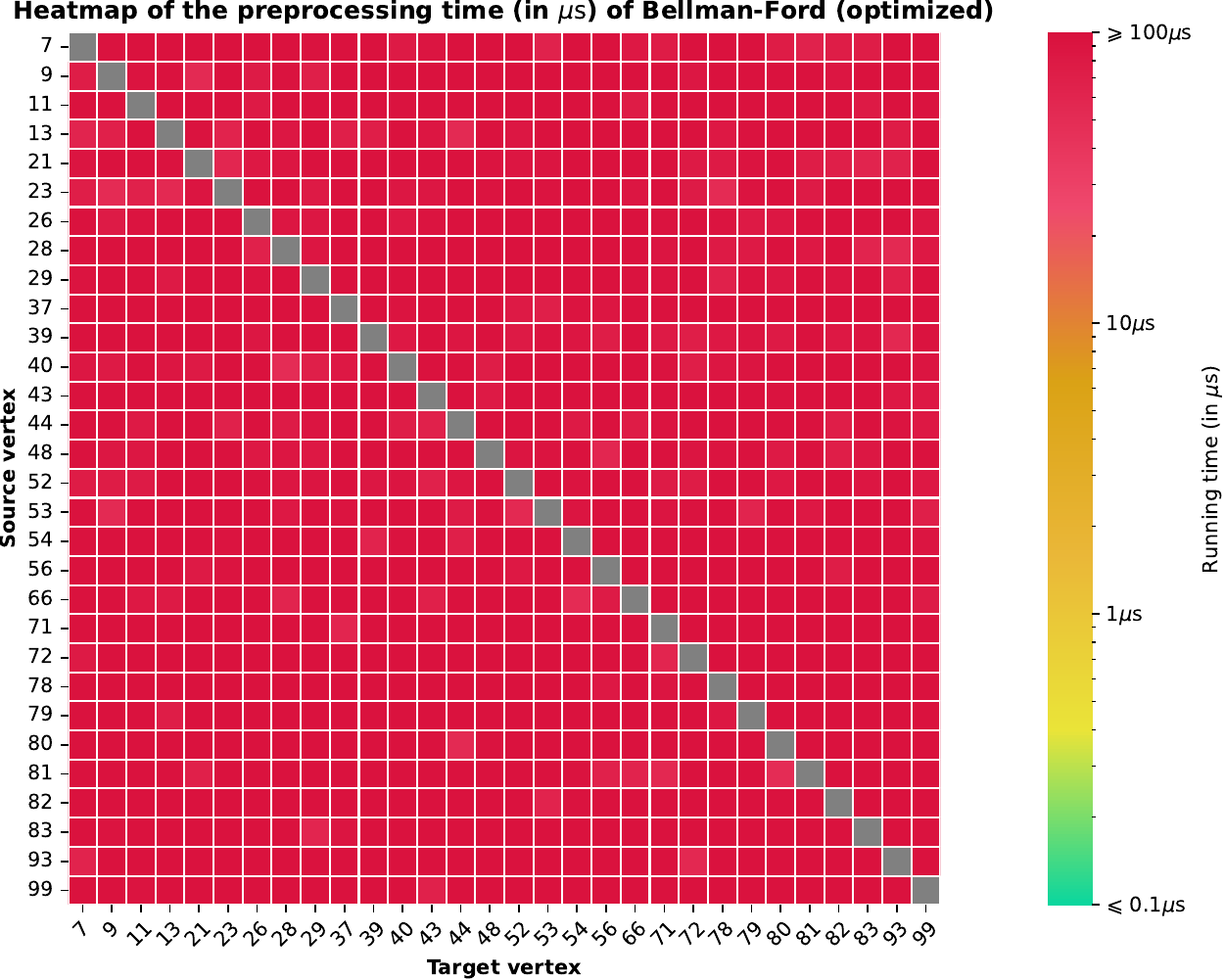}
            
            \vspace{0.3cm}
            
            \includegraphics[scale=0.65]{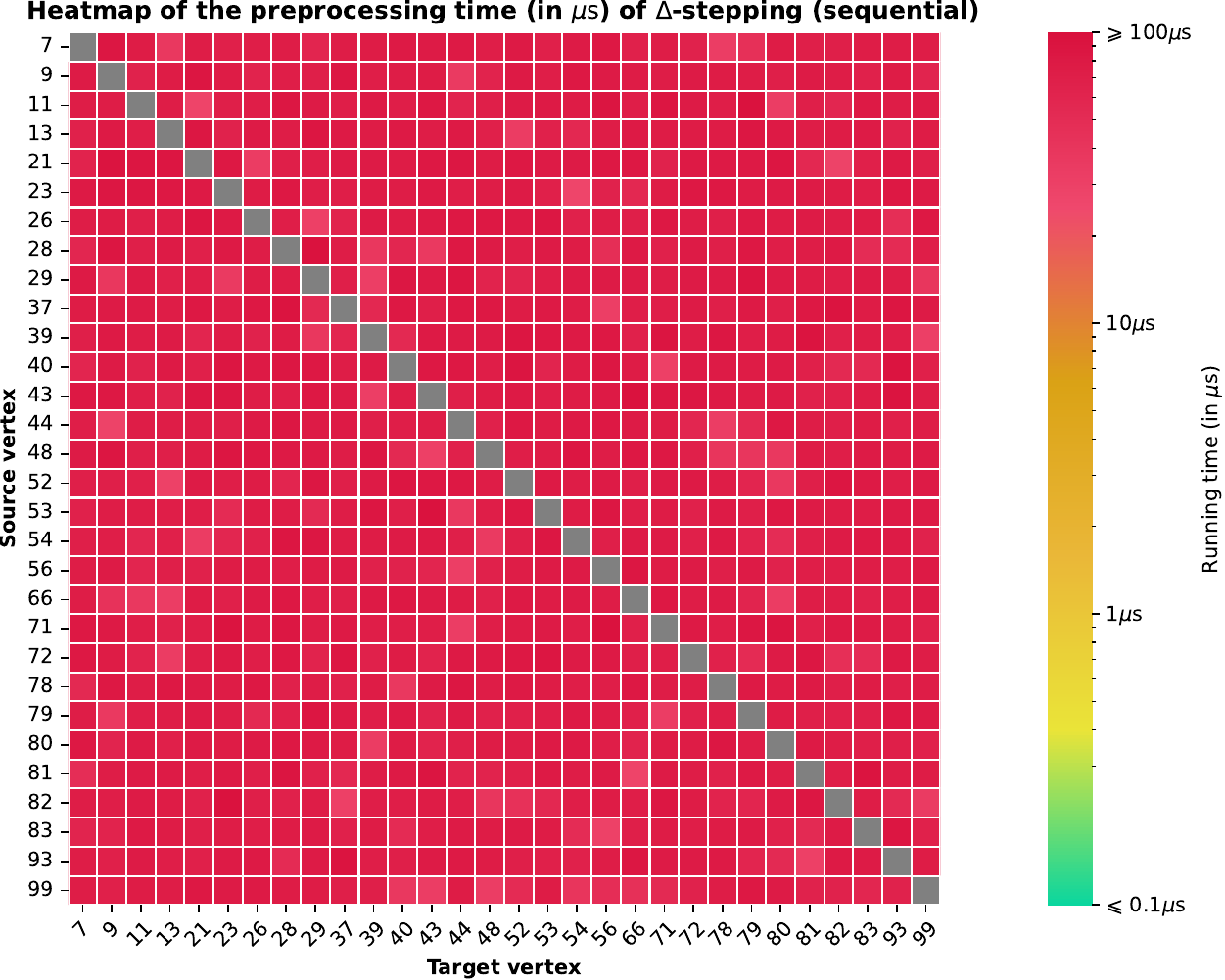}
            \captionof{figure}{\textit{Heatmaps of the pre-processing time for optimized \textsc{Bellman-Ford}'s and \textsc{$\Delta$-stepping} algorithm}.}
            \end{figure}
        \end{center}

         \begin{center}
         \begin{figure}[H]
            \includegraphics[scale=0.65]{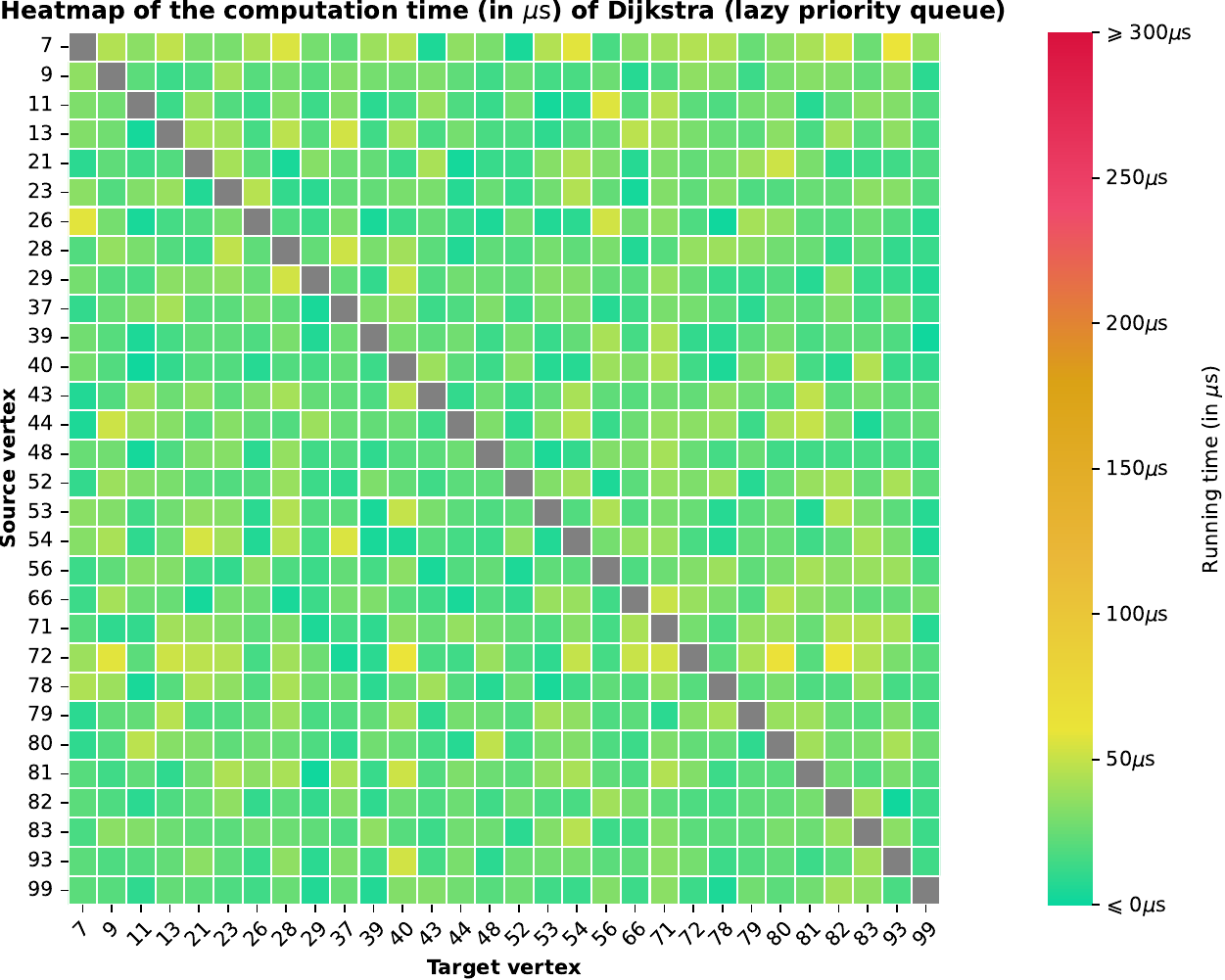}
            
            \vspace{0.3cm}
            
            \includegraphics[scale=0.65]{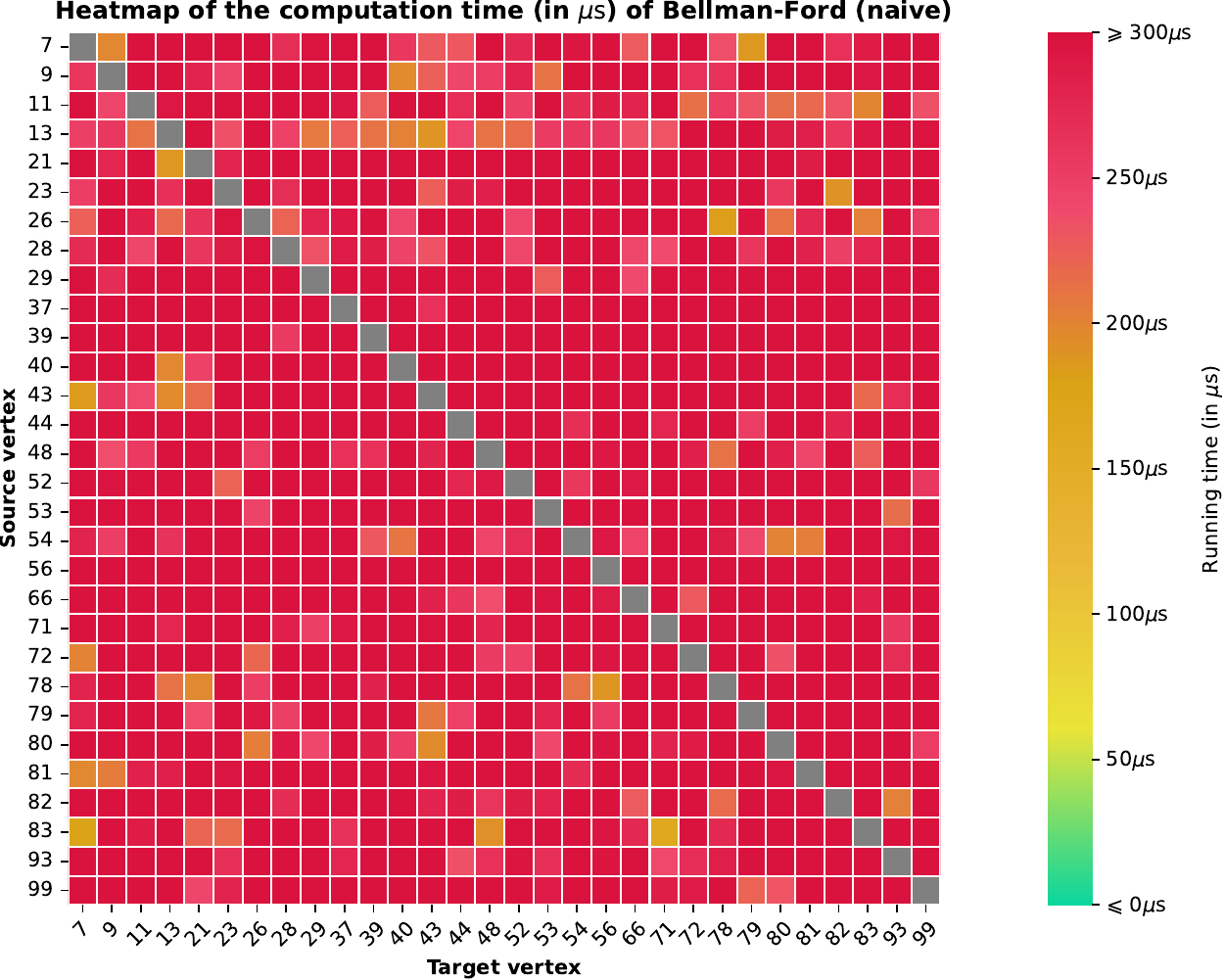}
            \captionof{figure}{\textit{Heatmaps of the computation time for \textsc{Dijkstra}'s and naive \textsc{Bellman-Ford} algorithm}.}
            \end{figure}
        \end{center}

         \begin{center}
         \begin{figure}[H]
            \includegraphics[scale=0.65]{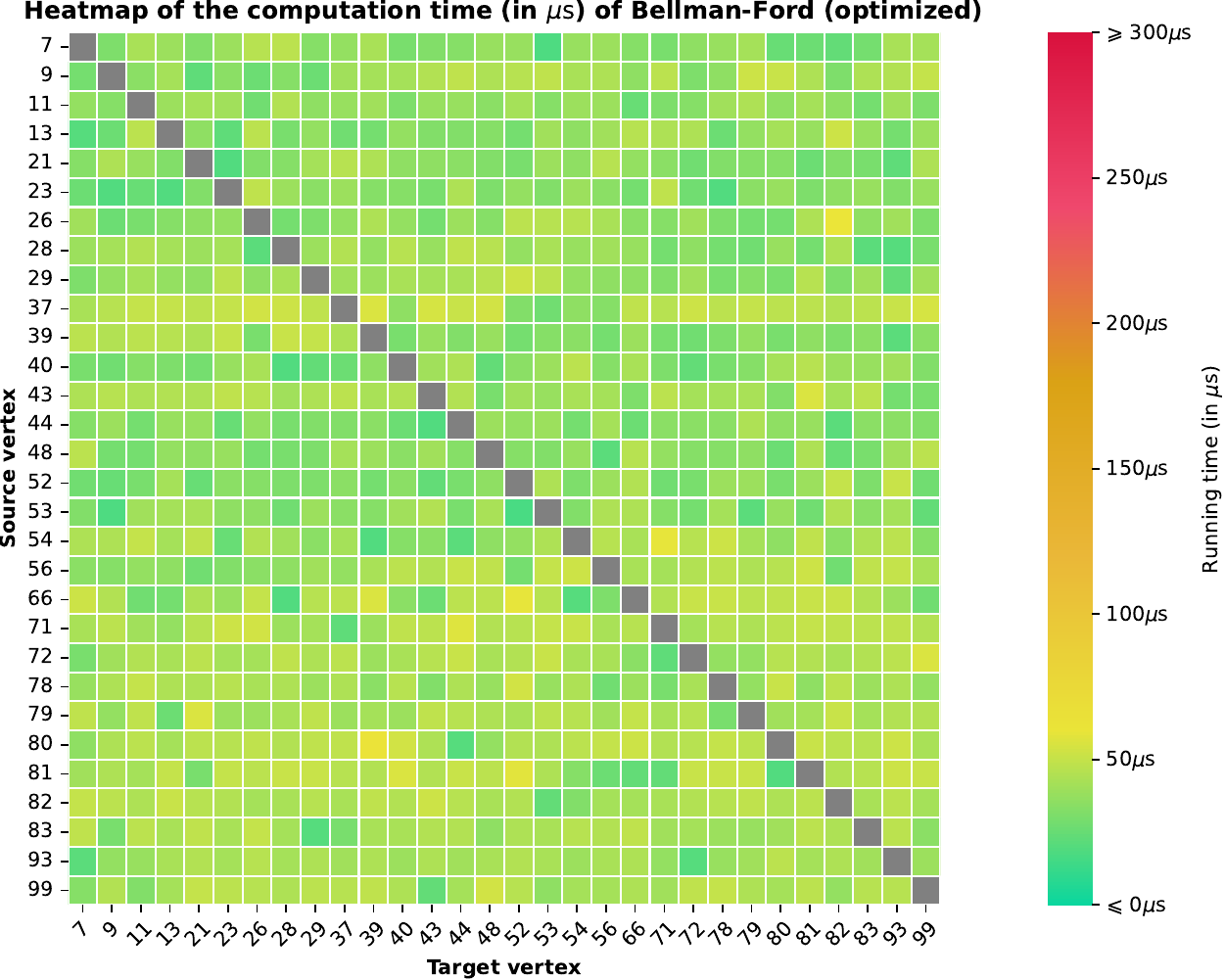}
            
            \vspace{0.3cm}
            
            \includegraphics[scale=0.65]{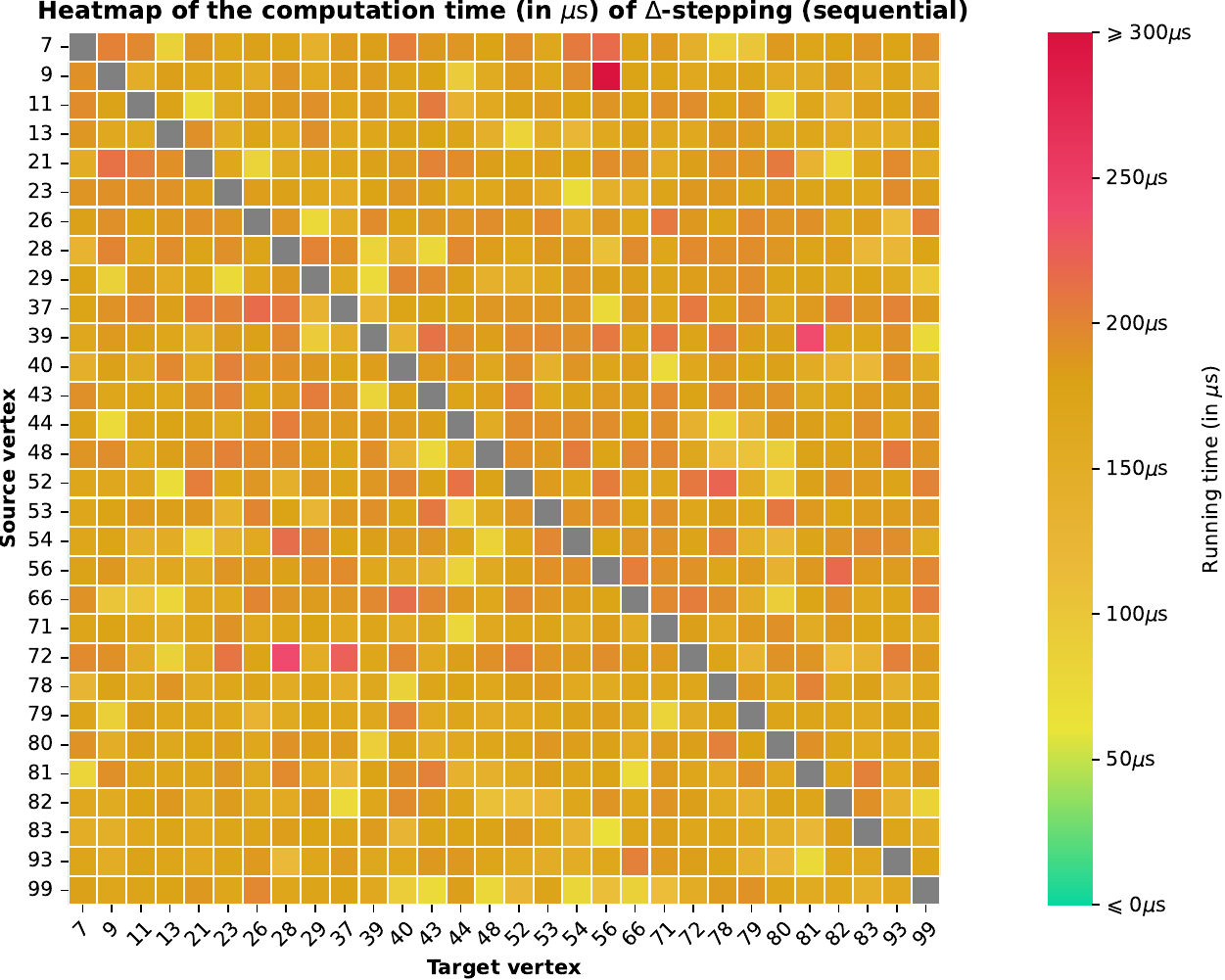}
            \captionof{figure}{\textit{Heatmaps of the computation time for optimized \textsc{Bellman-Ford}'s and \textsc{$\Delta$-stepping} algorithm}.}
            \end{figure}
        \end{center}
\vspace*{-2cm}
\section*{Annex B -- plots for task $2$}%

        \begin{center}
            \includegraphics[scale=0.6, angle = -90]{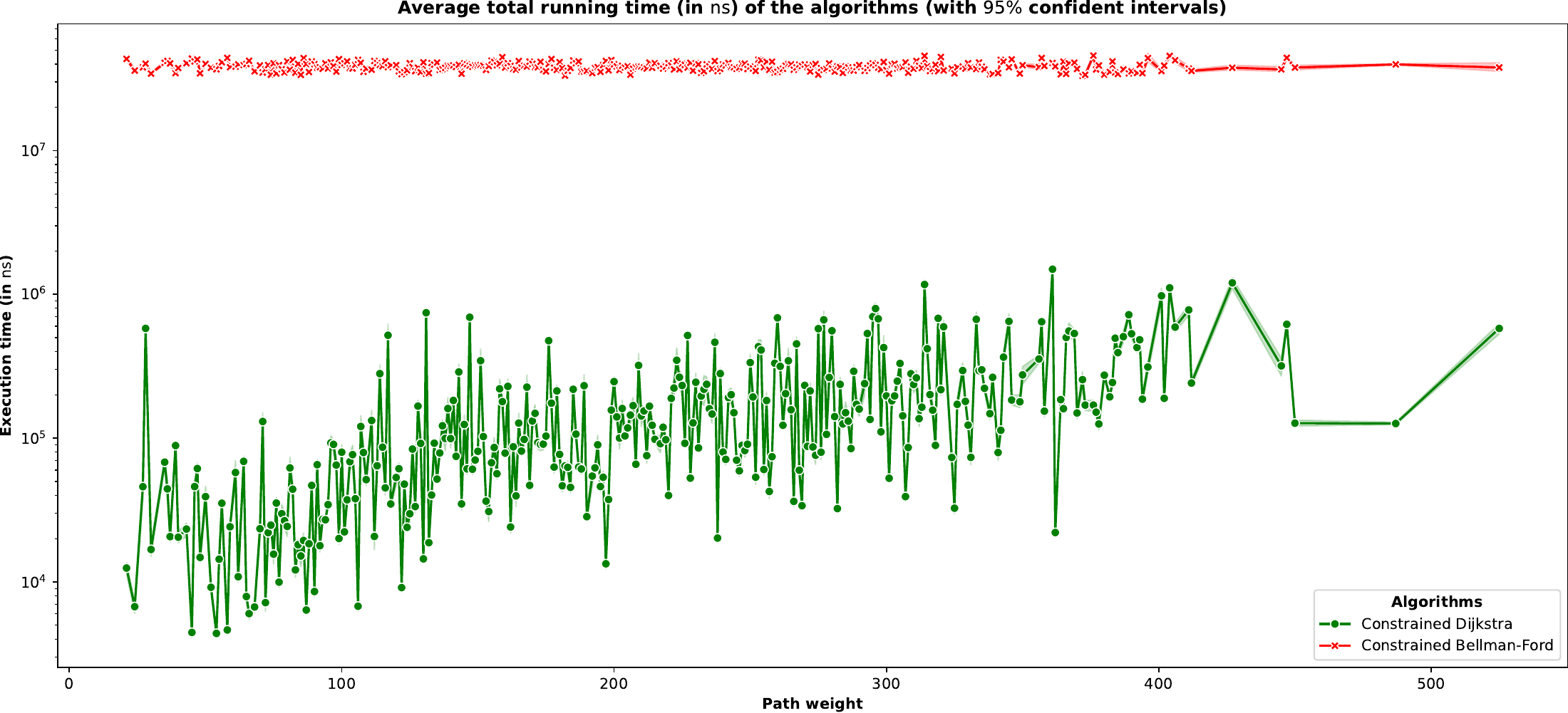}
            \captionof{figure}{\textit{Average total time (in $\mathrm{ns}$) of the two algorithms}.}
        \end{center}
        \clearpage

        \begin{center}
            \includegraphics[scale=0.6, angle = -90]{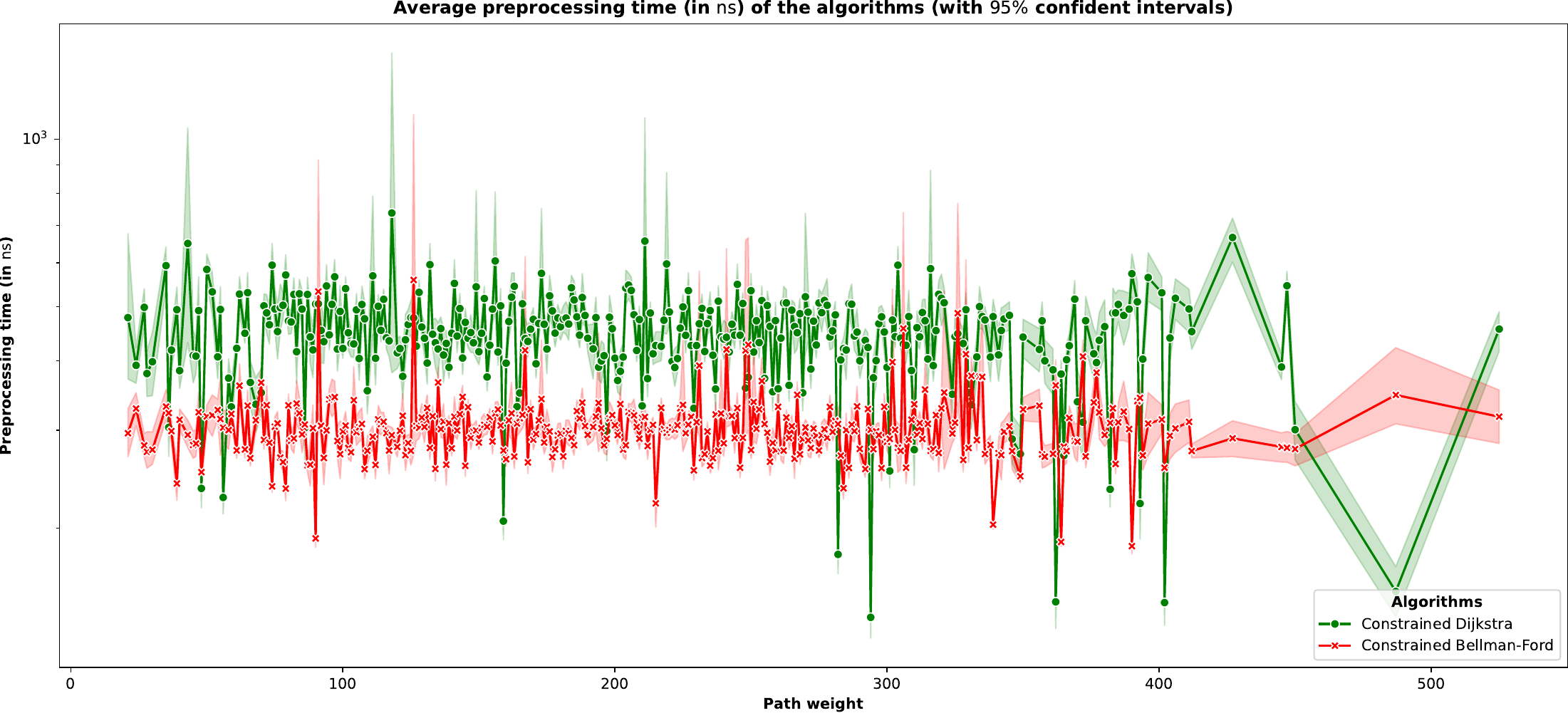}
            \captionof{figure}{\textit{Average pre-processing time (in $\mathrm{ns}$) of the two algorithms}.}
        \end{center}
        \clearpage

        \begin{center}
            \includegraphics[scale=0.6, angle = -90]{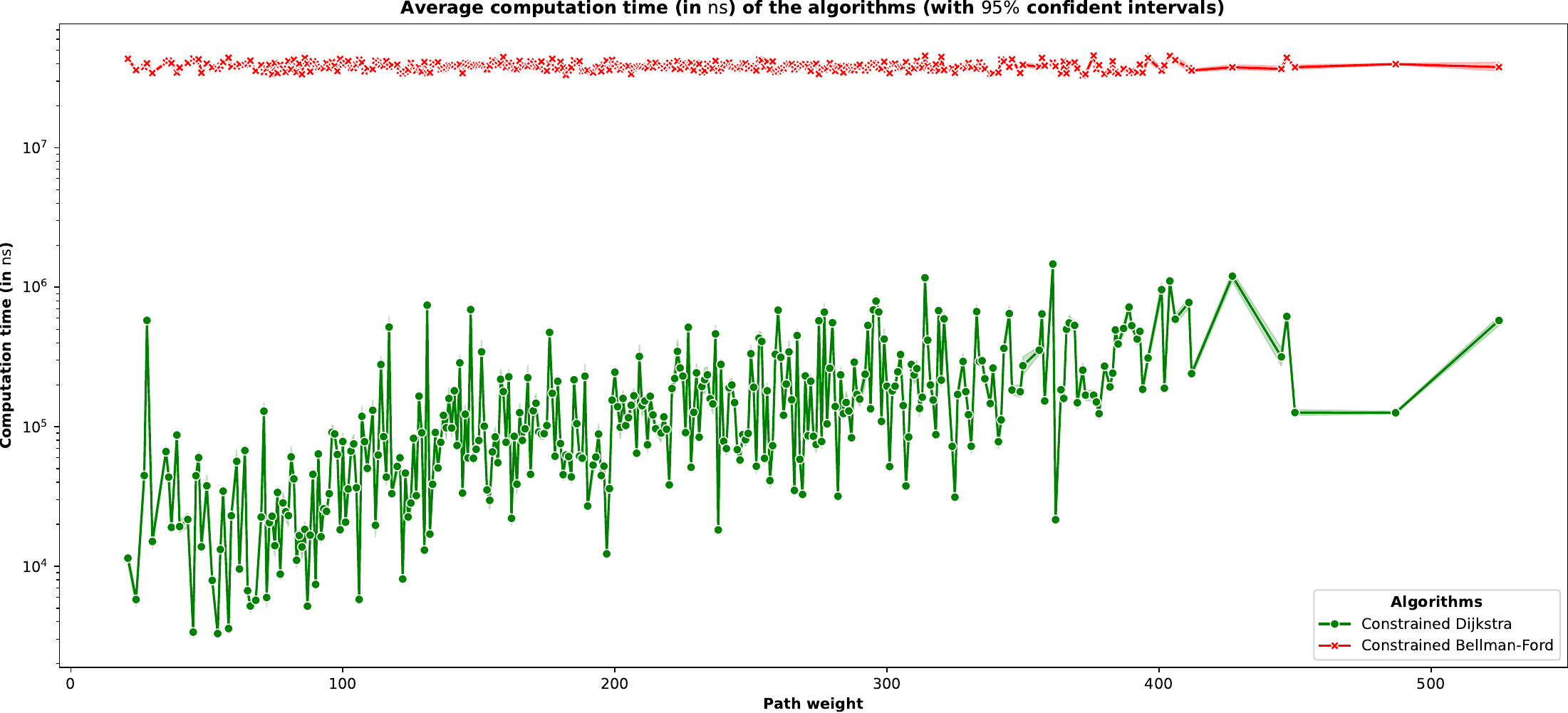}
            \captionof{figure}{\textit{Average computation time (in $\mathrm{ns}$) of the two algorithms}.}
        \end{center}
		\clearpage

        \begin{center}
         \begin{figure}[H]
            \includegraphics[scale=0.65]{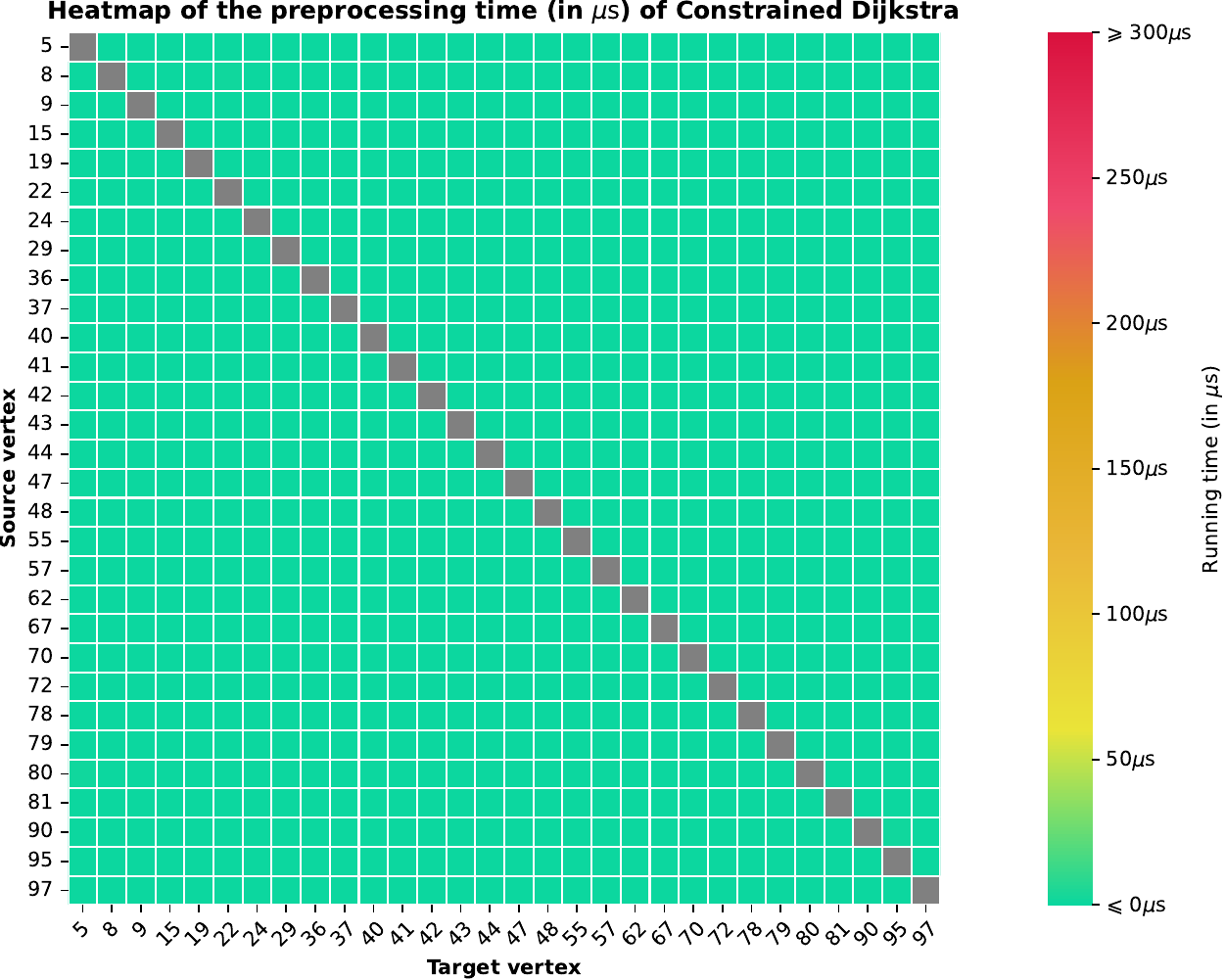}
            
            \includegraphics[scale=0.65]{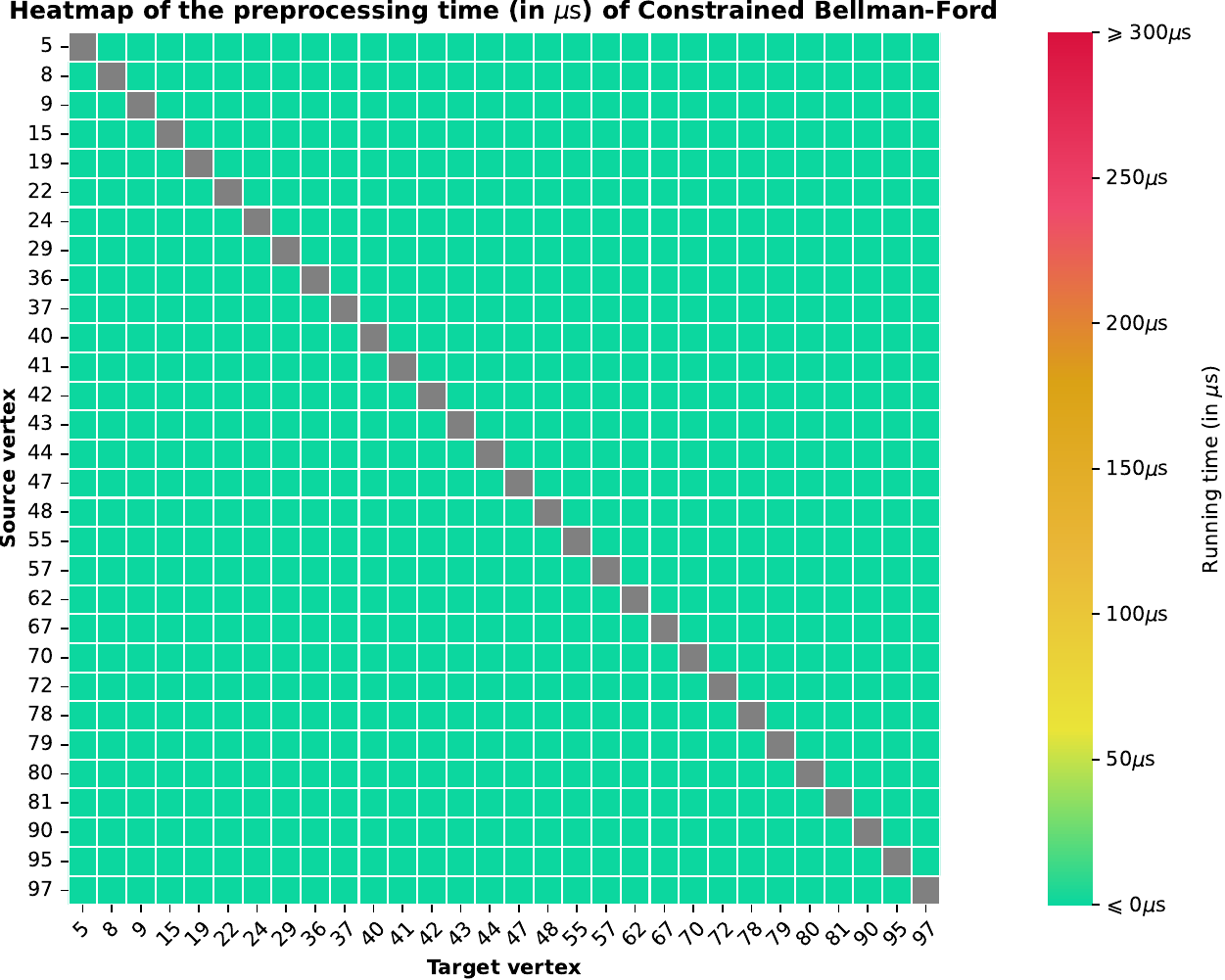}
            \captionof{figure}{\textit{Heatmaps of the preprocessing time for constrained \textsc{Dijkstra}'s and constrained \textsc{Bellman-Ford} algorithm}.}
            \end{figure}
        \end{center}
        \clearpage

        \begin{center}
         \begin{figure}[H]
            \includegraphics[scale=0.65]{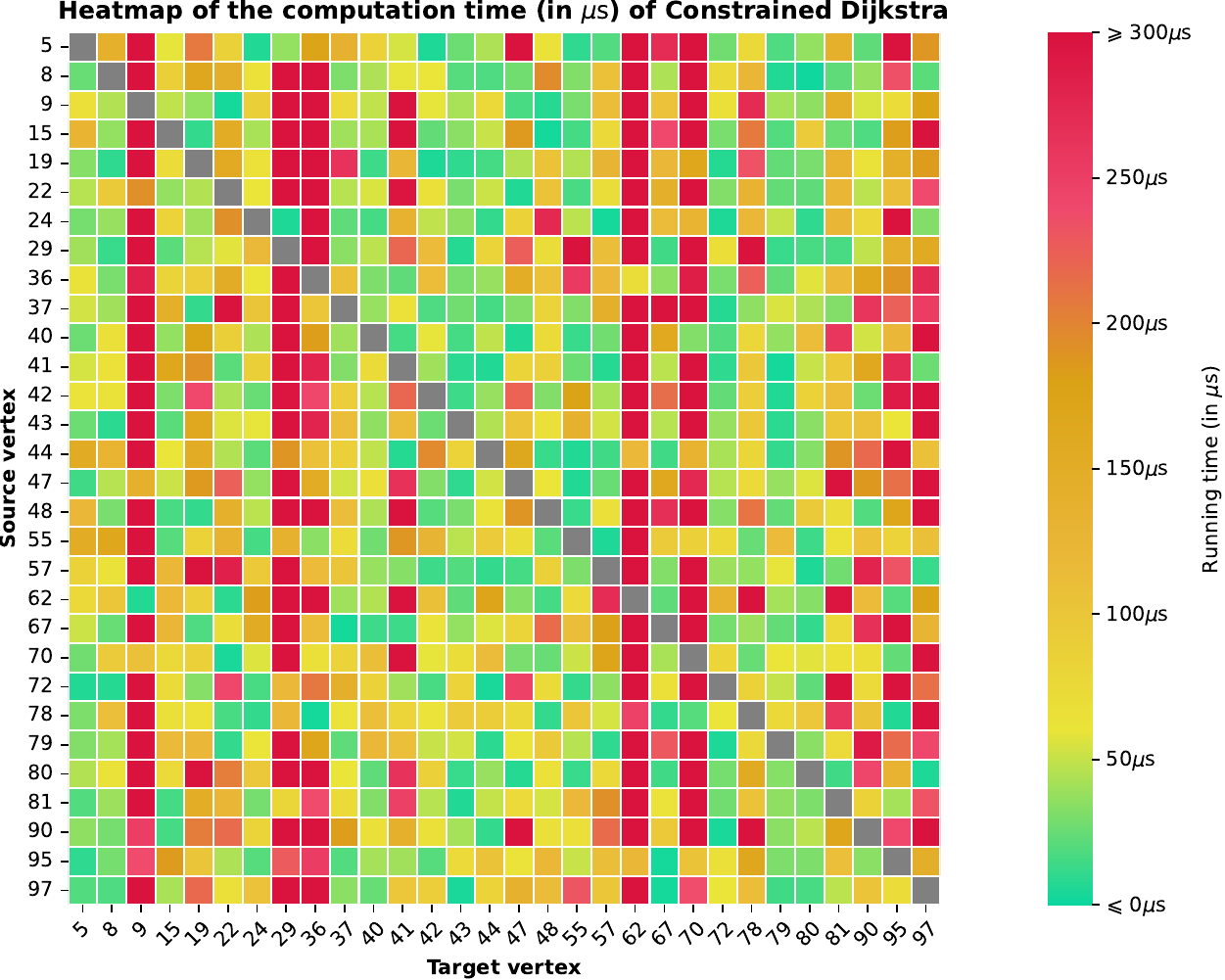}
            
            \vspace{0.3cm}
            
            \includegraphics[scale=0.65]{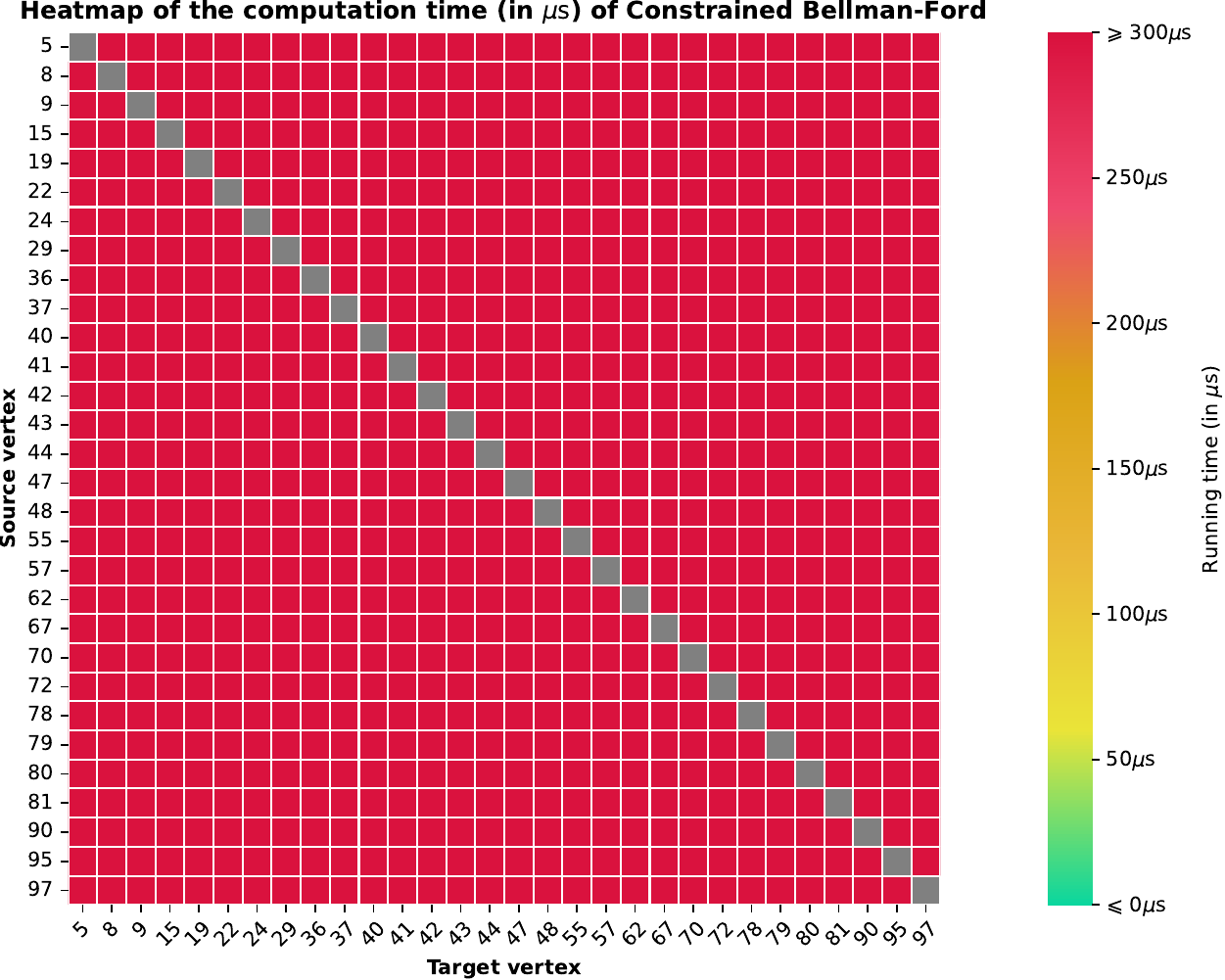}
            \captionof{figure}{\textit{Heatmaps of the computation time for constrained \textsc{Dijkstra}'s and constrained \textsc{Bellman-Ford} algorithm}.}
            \end{figure}
        \end{center}

        \begin{center}
         \begin{figure}[H]
            \includegraphics[scale=0.65]{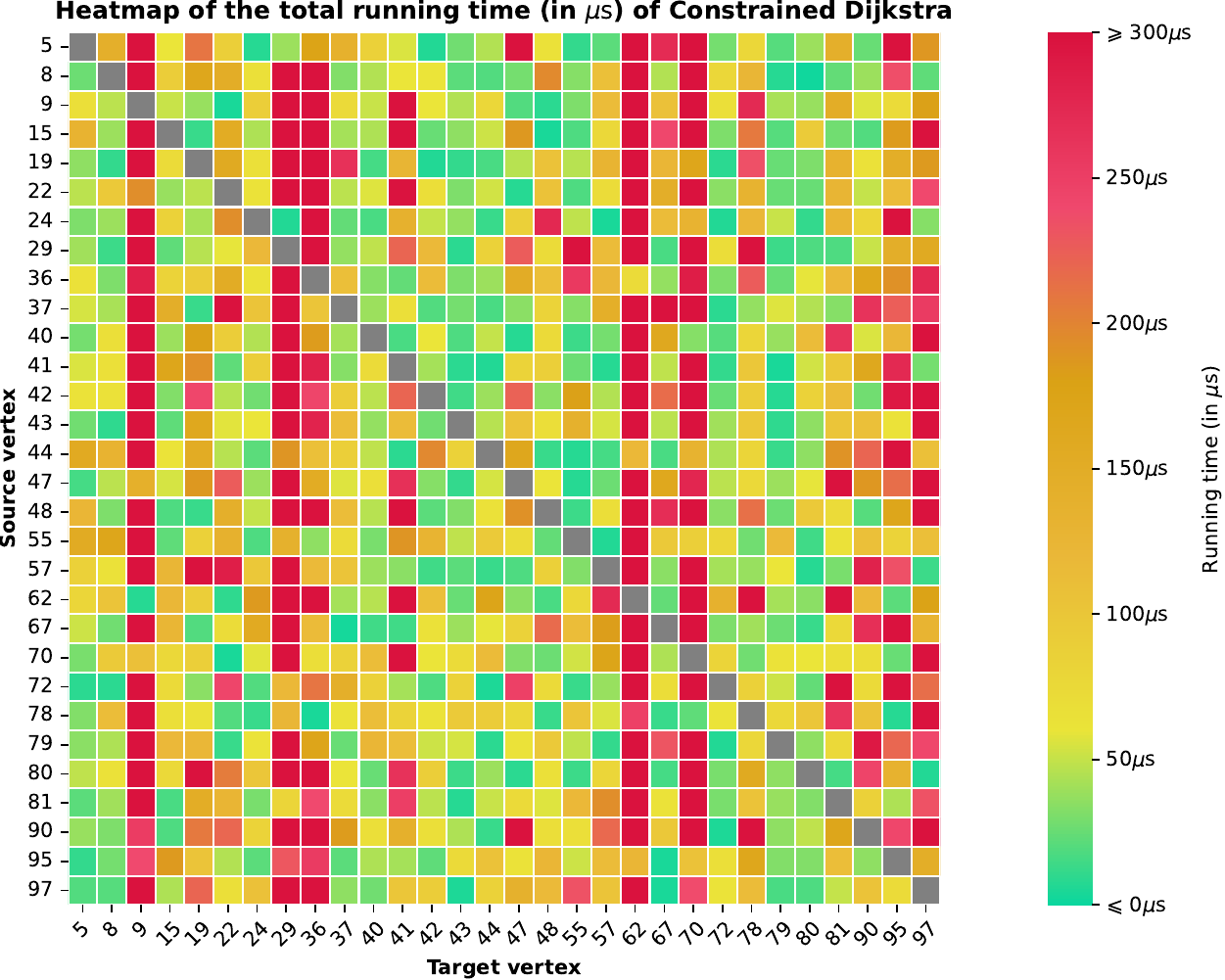}
            
            \vspace{0.3cm}
            
            \includegraphics[scale=0.65]{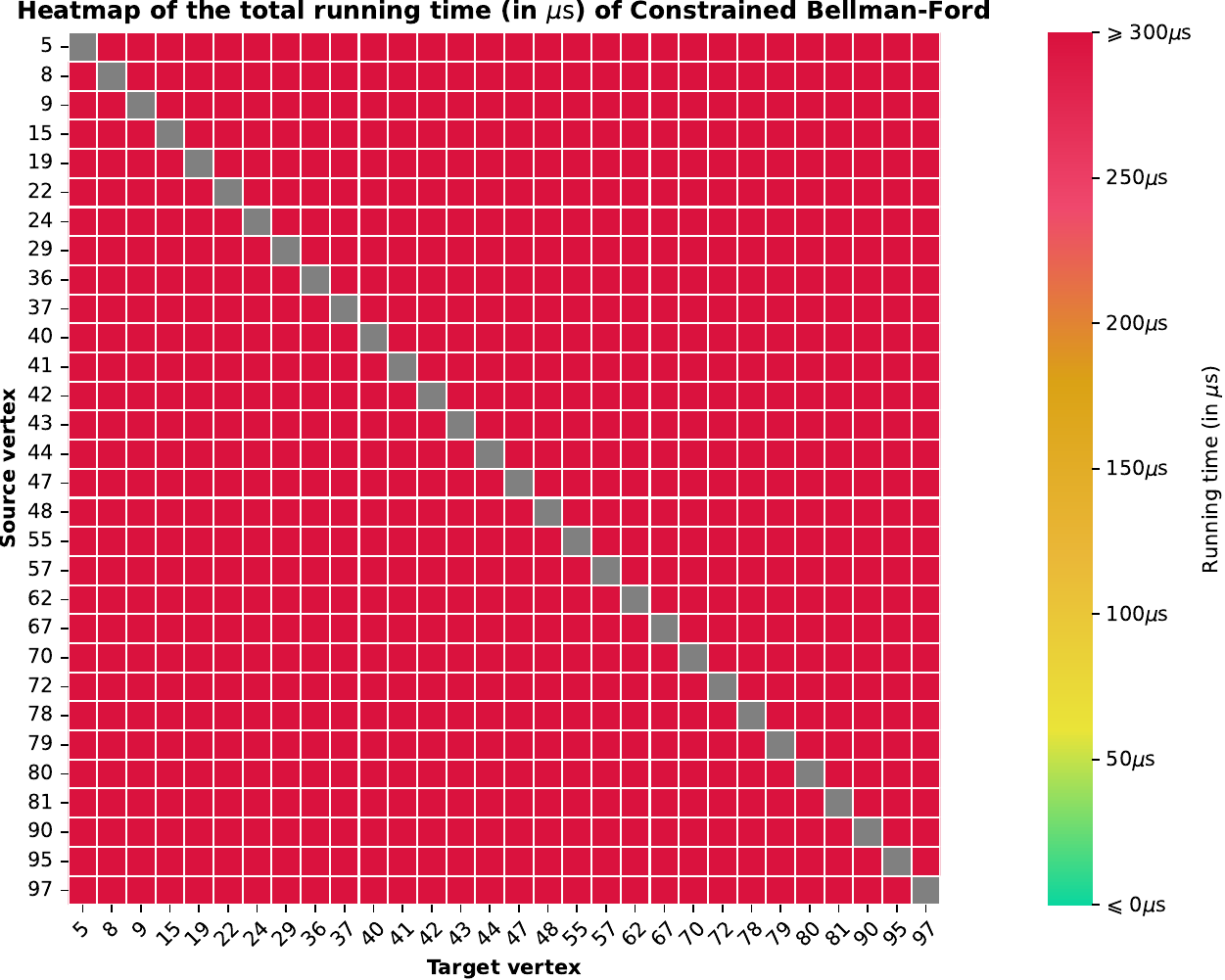}
            \captionof{figure}{\textit{Heatmaps of the total running time for constrained \textsc{Dijkstra}'s and constrained \textsc{Bellman-Ford} algorithm}.}
            \end{figure}
        \end{center}

\label{lastannex}%

\end{document}